\documentclass[12pt]{article}
\usepackage{amsmath,amssymb,amsthm}
\usepackage{txfonts,bm,pifont}
\usepackage[dvipdfmx]{graphicx}
\usepackage[pdfencoding=auto,bookmarks=true,bookmarksnumbered=true,colorlinks=true]{hyperref}
\usepackage{color}
\usepackage{float}
\usepackage{multirow,bigdelim}
\usepackage{comment}
\theoremstyle{plain}
\newtheorem{thm}{Theorem}[section]
\newtheorem{prop}[thm]{Proposition}
\newtheorem{defn}[thm]{Definition}
\newtheorem{lemma}[thm]{Lemma}

\newtheorem{rem}[thm]{Remark}


\definecolor{grey}{rgb}{0.5,0.5,0.5}

\makeatletter
\newdimen\LENB \newdimen\LENW \newdimen\THI 
\newdimen\LENWH \newdimen\LENTOT \newcount\N 
\def\vbrknlnele#1#2#3{
  \LENB=#1pt \LENW=#2pt \THI=#3pt
  \LENWH=\LENW \divide\LENWH by 2
  \LENTOT=\LENB \advance\LENTOT by \LENW
  \vbox to \LENTOT{
    \vbox to \LENWH{}
    \nointerlineskip
    \vbox to \LENB{\hbox to \THI{\vrule width \THI height \LENB}}
    \nointerlineskip
    \vbox to \LENWH{}
  }}

\def\vbrknln#1{
  \N=#1
  \vcenter{
    \vbox{
      \loop\ifnum\N>0
        \vbox to 4pt{\vbrknlnele{2}{2}{0.1}}
        \nointerlineskip
        \advance\N by -1
      \repeat
  }}}

\def\hbrknlnele#1#2#3{
  \LENB=#1pt \LENW=#2pt \THI=#3pt
  \LENTOT=\LENB \advance\LENTOT by \LENW
  \vcenter{
    \vbox to \THI{
      \hbox to \LENTOT{
        \hfil
        \vrule width \LENB height \THI
        \hfil}
  }}}

\def\hblele{\hbrknlnele{2}{2.2}{0.1}}

\def\hblfil{\cleaders\hbox{$ \m@th \mkern1mu \hblele \mkern1mu
$}\hfill} 
\makeatother
\begin{document}
\begin{center}
\begin{large}
Log-Aesthetic Curves: Similarity Geometry,\\ Integrable Discretization and Variational Principles\\[5mm]
\end{large}
\begin{normalsize}
Jun-ichi {\sc Inoguchi}\\
Institute of Mathematics, University of Tsukuba\\
Tsukuba 305-8571, Japan\\
e-mail: inoguchi@math.tsukuba.ac.jp\\[2mm]
Yoshiki {\sc Jikumaru}\\
Institute of Mathematics for Industry, Kyushu University\\
744 Motooka, Fukuoka 819-0395, Japan\\
e-mail: y-jikumaru@imi.kyushu-u.ac.jp\\[2mm]
Kenji {\sc Kajiwara}\\
Institute of Mathematics for Industry, Kyushu University\\
744 Motooka, Fukuoka 819-0395, Japan\\
e-mail: kaji@imi.kyushu-u.ac.jp\\[2mm]
Kenjiro T. {\sc Miura}\\
Graduate School of Science and Technology, Shizuoka University\\
3-5-1 Johoku, Hamamatsu, Shizuoka, 432-8561, Japan\\
e-mail: miura.kenjiro@shizuoka.ac.jp\\[2mm]
Wolfgang K. {\sc Schief}\\
School of Mathematics and Statistics, The University of New South Wales\\
Sydney, NSW 2052, Australia\\
e-mail: w.schief@unsw.edu.au
\end{normalsize}
\end{center}
\begin{abstract}
In this paper, we consider a class of plane curves called log-aesthetic curves and their
generalization which are used in computer aided geometric design. We consider these curves in the
framework of the similarity geometry and characterize them as invariant curves under the integrable
flow on plane curves which is governed by the Burgers equation. We propose a variational principle
for these curves, leading to the stationary Burgers equation as the Euler-Lagrange equation. As an
application of the formulation developed here, we propose a discretization of these curves and the
associated variational principle which preserves the underlying integrable structure. We finally
present algorithms for the generation of discrete log-aesthetic curves for given ${\rm G}^1$ data
based on the similarity geometry.  Our method is able to generate $S$-shaped discrete curves with an
inflection as well as $C$-shaped curves according to the boundary condition.  The resulting discrete
curves are regarded as self-adaptive discretization and thus high-quality even with a small number
of points.
\end{abstract}
\section{Introduction}
\noindent In this paper, we consider a class of plane curves in computer aided geometric design
called the {\em log-aesthetic curves} (LAC) and their generalization called the {\em quasi aesthetic
curves} (qAC), and present a new mathematical characterization based on the theory of integrable
systems and similarity geometry. We then construct the discrete analogue of LAC and qAC within the
above-mentioned framework, which gives a new implementation of LAC and qAC with a sound mathematical
background as discrete curves.

In the previous paper \cite{IKMSSS_CAGD:2018}, we have announced the similarity geometric framework
of the LAC and the qAC, where these curves have been characterized as the invariant curves
under the integrable flow on the plane curves preserving the similarity arc length. More precisely,
the evolution of the curves is governed by the similarity curvature which is characterized by
the stationary solutions of integrable nonlinear partial differential equation arising from the
geometric setting.  In addition, we have introduced a fairing energy functional and formulated
the LAC and the qAC in terms of a variational principle. Here, we first present a
detailed account of those results.

Secondly, we construct a discrete analogue of the LAC and the qAC based on the
above formulation, where these curves are characterized as the invariant discrete curves under
the discrete integrable flow on discrete plane curves preserving the similarity arc length. We then
introduce a discrete fairing functional and formulate these discrete curves in terms of a discrete
variational principle. The discrete curves obtained in this manner are not na\"ive approximations of
the original LAC and qAC but admit their own natural geometric characterization.

Finally, we give an implementation of generation method of the discrete LAC obtained above for
given endpoints and associated tangent vectors. We consider the cases of the discrete LAC
without/with an inflection (``$C$-shaped'' / ``$S$-shaped''). We note that the discrete LAC based on
the similarity geometry gives a kind of self-adaptive mesh discretization of the LAC, since
we have dense points where the curvature is large and coarse points where the curvature is small.

%
\section{Log-aesthetic curves and similarity geometry}
Originally, LAC has been studied in the framework of Euclidean geometry. Before proceeding to
LAC, we give a brief account of the treatment of plane curves in Euclidean geometry. Let
$\gamma(s)\in\mathbb{R}^2$ be an arc length parametrized plane curve and $s$ be arc length.  We
introduce the Frenet frame $F^{\rm E}(s)\in {\rm SO(2)}$ by
\begin{equation}
F^{\rm E}(s) = (T^{\rm E}(s), N^{\rm E}(s)),\quad  
\frac{d\gamma(s)}{ds} = T^{\rm E}(s),\quad N^{\rm E}(s) = JT^{\rm E}(s),
\end{equation}
where $T^{\rm E}(s)$ and $N^{\rm E}(s)$ are the tangent and the normal vector fields, respectively,
and $J$ is the positive $\pi/2$-rotation. Since $|T^{\rm E}(s)|=1$ by definition of arc length,
we may write $T^{\rm E}(s)={}^t(\cos\theta(s), \sin\theta(s))$, where $\theta$ is called the {\em
angle function}.  The Frenet frame satisfies the {\em Frenet formula}
\begin{equation}
 \frac{dF^{\rm E}(s)}{ds} = F(s)L^{\rm E}(s),\quad 
L^{\rm E}(s) = \left(\begin{array}{cc} 0 & -\kappa(s) \\ \kappa(s) & 0 \end{array}\right),\label{eqn:Frenet_formula}
\end{equation}
where $\kappa(s)$ is the curvature. Note that the curvature $\kappa$ is related to 
the signed radius of curvature $q(s)$ and the angle function $\theta(s)$ by
\begin{equation}
 q(s) = \frac{1}{\kappa(s)},\quad \kappa(s) = \frac{d\theta(s)}{ds}.\label{eqn:q_kappa}
\end{equation}

According to \cite{Miura}, an arc length parametrized plane curve $\gamma(s)\in\mathbb{R}^2$ is said
to be a LAC of slope $\alpha$ if its signed curvature radius $q$ satisfies
\begin{equation} \label{eqn:LAC_curvature}
 q(s)^{\alpha}= c_0s+c_1\quad (\alpha\not=0),\quad q(s)=\exp(c_0s+c_1)
\quad (\alpha=0),\quad c_0, c_1\in\mathbb{R}.
\end{equation}
Originally, LAC is characterized as a family of arc length parametrized curves whose {\em
logarithmic curvature histogram}, a graph of $\log|q|$ versus $ \log\left|\frac{ds}{d(\log
|q|)}\right|$, is a line, and $\alpha$ is its slope \cite{Harada:LAC1,Harada:LAC2}.

\begin{figure}[t]
\begin{center}
\includegraphics[width=0.8\linewidth]{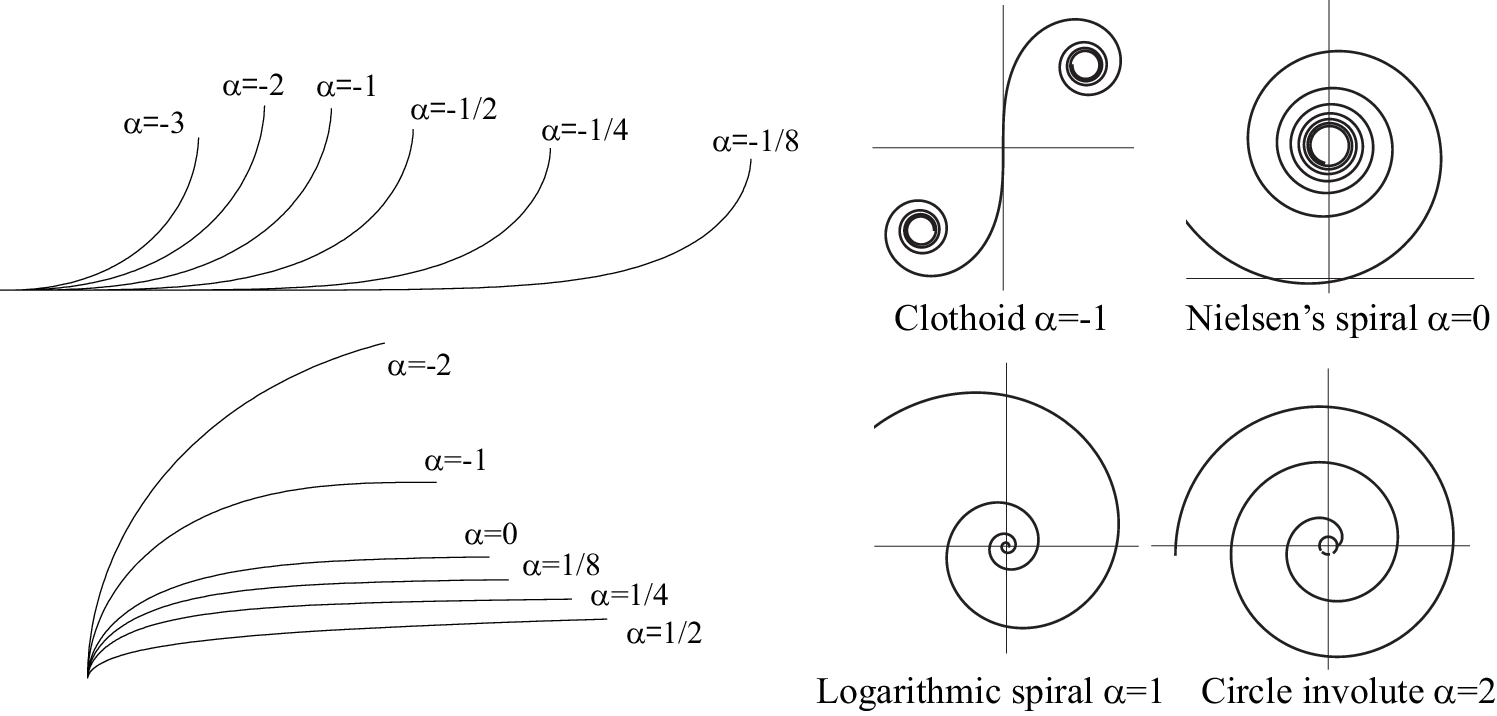} 
\end{center}
\caption{Log-aesthetic curves. Left: LAC for various parameters. Right: LAC for $\alpha=-1,0,1,2$.}\label{fig:LAC}
\end{figure}

The class of LAC includes some well known plane curves. For instance, the logarithmic spiral, the
clothoid, the Nielsen spiral are included as LAC of slope $1$, $-1$, and $0$, respectively.  The LAC
of slope $2$ is also known as the circle involute curve. These examples are illustrated in
Figure \ref{fig:LAC}.

\begin{figure}[hbtp]
\centering
\includegraphics[width=0.8\linewidth]{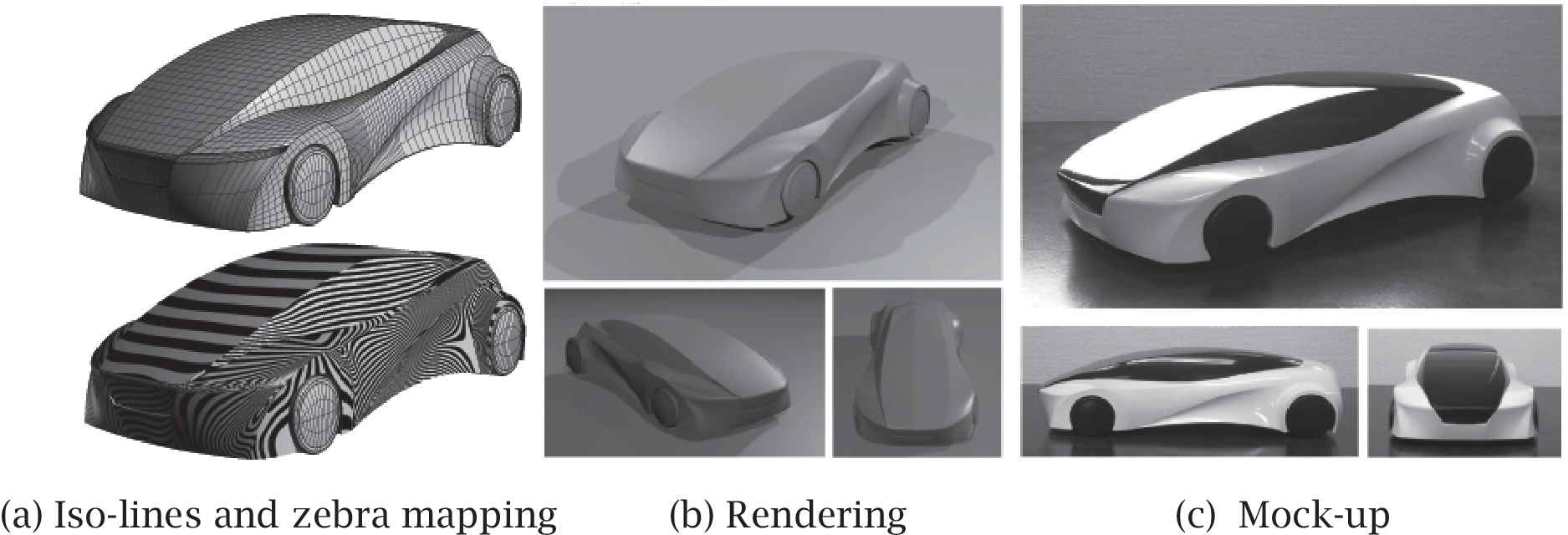}
\caption{A car model designed by means of LA splines and its mock-up.}
\label{fig:Zebra0}
\end{figure}

LAC are now maturing in industrial and graphics design practices. Figure \ref{fig:Zebra0} shows the
practical example of a car designed using LA splines. Figure \ref{fig:Zebra0}(a) shows free-form
surface iso-parametric lines generated using LA splines and corresponding zebra maps. Figure
\ref{fig:Zebra0}(b) shows the geometric model with a special lighting condition and
\ref{fig:Zebra0}(c) are photos of a manufactured mockup based on the geometric model. Note that the
roof of the car is designed by an LA spline curve with three segments and its zebra maps indicate
that the surface is of high quality. Based on our experience, LA splines are generated with most
$G^2$ Hermite data. Another direction of application is developed in architecture design
\cite{Suzuki:LAC}. For more details of the LAC, we refer to \cite{IZM:2019,Miura2014,Miura2016}.
%
%

Those studies have been carried out based on the basic characterization \eqref{eqn:LAC_curvature} in
the framework of Euclidean geometry. However, \eqref{eqn:LAC_curvature} is too simple to identify
the underlying geometric structure. Consequently, we do not have a good guideline as to how to
generate a larger class of aesthetic geometric objects including LAC based on a sound mathematical
background.  As we have announced in the previous paper \cite{IKMSSS_CAGD:2018}, it is natural to
adopt the framework of similarity geometry, which is a Klein plane geometry associated with the
group of similarity transformations, \textit{i.e.}, isometries and scalings:
\begin{displaymath}
\mathbb{R}^2\ni \bm{p}\longmapsto rA\bm{p}+\bm{b},\quad A\in{\rm SO}(2),\quad r\in\mathbb{R}_{+},
\quad \bm{b}\in\mathbb{R}^2.
\end{displaymath}
The natural parameter of plane curves in similarity geometry is the angle function
$\theta=\int \kappa(s)\>ds$. Let $\gamma(\theta)\in\mathbb{R}^2$ be a plane curve in similarity geometry
parametrized by $\theta$.  
We introduce the similarity Frenet frame $F(\theta)$ by
\begin{equation}
 F(\theta) = (T,N)\in{\rm CO}^{+}(2) = \{rA\,|\, r\in\mathbb{R}_{+},\ A\in{\rm SO}(2) \},
\end{equation}
where 
\begin{equation}\label{eqn:similarity_tangent}
T=\frac{\mathrm{d}\gamma}{\mathrm{d}\theta},
\quad
N=J\frac{\mathrm{d}\gamma}{\mathrm{d}\theta}, 
\end{equation}
are the similarity tangent and normal vector fields, respectively.
\begin{figure}[t]
\begin{center}
\includegraphics[scale=0.8]{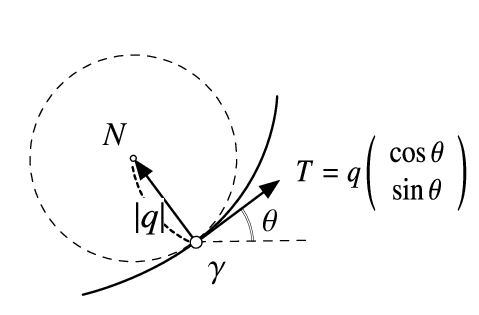}
\end{center}
\caption{Description of plane curves in similarity geometry.}\label{fig:similarity_frame}
\end{figure}
Note that $|T(\theta)|=|q|$ which follows from $|T^{\rm E}(s)|=1$ and $d\theta/ds = \kappa = 1/q$
(see Figure \ref{fig:similarity_frame}). Then, the (Euclidean) Frenet formula
\eqref{eqn:Frenet_formula} implies that the similarity Frenet frame satisfies the {\em similarity
Frenet formula}
\begin{equation}\label{eqn:similarity_Frenet}
 \frac{dF}{d\theta} = FL,\quad L=\left(\begin{array}{cc} -u& -1\\1 & -u \end{array}\right),
\end{equation}
for some function $u(\theta)$ which is called the \emph{similarity curvature}.  
Moreover, the similarity curvature $u$ is related to the signed
curvature radius $q$ by the \textit{Cole-Hopf transformation}:
\begin{equation}\label{eqn:C-H} 
u:=-\frac{1}{q}\>\frac{\mathrm{d}q}{\mathrm{d}\theta}. 
\end{equation}
One can check that a plane curve in similarity geometry is uniquely determined by the similarity
curvature up to similarity transformations.

The notion of LAC may be shown to be invariant under the similarity transformations. For instance,
the slope $\alpha$ is expressed as $\alpha = 1+(1/u^2)\,du/d\theta.$ In other words, the LAC is
reformulated in terms of the similarity geometry as follows \cite{Inoguchi, SS1}. Due to
\eqref{eqn:q_kappa} and \eqref{eqn:LAC_curvature}, a plane curve $\gamma(\theta)$ in similarity
geometry is a LAC of slope $\alpha$ if its similarity curvature satisfies the Bernoulli equation
\begin{equation}\label{LAC}
\frac{\mathrm{d}u}{\mathrm{d}\theta}=(\alpha-1)u^2,
\end{equation}
so that the similarity curvature itself is explicitly given by
\begin{equation}\label{eqn:LAC_similarity_curvature}
 u = -\frac{\lambda}{(\alpha-1)\lambda\theta+1},\quad \lambda\in\mathbb{R}.
\end{equation}
%
Based on this reformulation, the qAC is introduced in the following manner \cite{SS2}. A plane curve
in similarity geometry is a qAC of slope $\alpha$ if its similarity curvature obeys the Riccati
equation
\begin{equation}
\frac{\mathrm{d}u}{\mathrm{d}\theta}=(\alpha-1)u^2+c,\quad c\in\mathbb{R}.
\label{eqn:Riccati_LAC}
\end{equation}
%
\section{Burgers flow on similarity plane curves}
One of the key techniques to understand the LAC and the qAC is to consider the integrable (time)
evolution of plane curves that preserves the invariant parameter of the similarity geometry, which
is known to be described by the Burgers hierarchy \cite{CQ}.  The simplest evolution is given by
\begin{equation}\label{CQflow}
\frac{\partial}{\partial t}\gamma=(b-u)T-N,\quad b\in\mathbb{R},
\end{equation}
which is rewritten in terms of the similarity Frenet frame $F(\theta)$ as
\begin{equation}\label{eqn:similarity_frame}
\frac{\partial F}{\partial t}
=FM,\quad M = 
\left(\begin{array}{cc}
-\frac{\partial u}{\partial \theta}+u^2+1-bu & -b\\
b & -\frac{\partial u}{\partial \theta}+u^2+1-bu
\end{array}\right),
\end{equation}
The compatibility condition 
$\partial L/\partial t - \partial M/\partial \theta = LM-ML$
of \eqref{eqn:similarity_Frenet} and \eqref{eqn:similarity_frame}
yields
the \textit{Burgers equation}
\begin{equation}
\frac{\partial u}{\partial t}=
\frac{\partial}{\partial \theta}\left(
\frac{\partial u}{\partial \theta}- u^2 +b u\right).  \label{eqn:Burgers}
\end{equation}
Therefore, the evolution \eqref{CQflow} is referred to as the {\em Burgers flow}.  The Burgers
equation is linearized in terms of the signed curvature radius via the Cole-Hopf transformation
\eqref{eqn:C-H} according to
\begin{displaymath}
\frac{\partial q}{\partial t}=\frac{\partial^2 q}{\partial \theta^2}+
b\frac{\partial q}{\partial \theta}. 
\end{displaymath}
Imposing the \textit{stationarity ansatz} $\partial u/\partial t=0$ reduces the Burgers equation
\eqref{eqn:Burgers} to the Riccati equation
\begin{equation}\label{St-Burgers}
\frac{\partial u}{\partial \theta}=u^2-bu+c,\quad c\in\mathbb{R}.
\end{equation}
In particular, putting $b=0$, we recover the Riccati equation \eqref{eqn:Riccati_LAC} with
$\alpha=2$. We note that \eqref{eqn:Riccati_LAC} is obtained formally from \eqref{St-Burgers} by
making the substitution $u\to (\alpha-1)u$. In this sense, the qAC are characterized by the
stationary Burgers flow.  We also note that the parameter $b$ corresponds to a rotation of the
curve.
%
\section{Fairing energy in similarity geometry}
In this section, we present the details of a variational formulation of LAC and qAC.  To this end,
we introduce the \textit{fairing energy functional} $\mathcal{F}^{\lambda,a}$
\cite{IKMSSS_CAGD:2018}
\begin{equation}\label{eqn:fairing_functional}
\mathcal{F}^{\lambda,a}(\gamma)=\int_{\theta_1}^{\theta_2}
\frac{1}{2}
\left\{
a^2 u(\theta)^2+\lambda\left(\frac{q_1\,q_2}{q(\theta)^{2}}\right)^a
\right\}
\,\mathrm{d}\theta,
\end{equation}
where $a=\alpha-1$, $\lambda$ is an arbitrary constant and $q_i = q(\theta_i)$ ($i=1, 2$).  The
above functional is invariant under the similarity transformations and its name ``fairing energy''
is motivated by the fairing procedure in digital style design of industrial products.  To compute
the variation, we consider a deformation of $\gamma$ parametrized by
\begin{equation}\label{eqn:variation_gamma}
\overline{\gamma}= \gamma + \epsilon\delta\gamma,\quad 
\delta\gamma=\xi(\theta)T(\theta) + \eta(\theta)N(\theta),
\end{equation}
where $\delta\gamma$ is the variation of $\gamma$. We distinguish the quantities relevant to the
deformed curve from their undeformed counterparts by adding an overbar
$\overline{\phantom{\gamma}}$. For example, the angle function and the similarity curvature of
$\overline{\gamma}$ are denoted by $\overline{\theta}$ and $\overline{u}$, respectively. In order to
obtain the variation of $u$ and $\theta$ from \eqref{eqn:variation_gamma}, we first compare the
similarity Frenet formula for $\gamma$ and $\overline\gamma$:
\begin{equation}\label{eqn:gammabar_TN}
\begin{split}
& \frac{d\overline\gamma}{d\theta} 
= \frac{d\overline\gamma}{d\overline{\theta}}\,  \frac{d\overline\theta}{d\theta}
= (1+\epsilon\phi)T + \epsilon\psi N, \\[2mm]
& \phi(\theta) = \frac{d\xi}{d\theta} - u\xi -\eta,\quad
\psi(\theta) = \frac{d\eta}{d\theta} - u\eta + \xi,
\end{split}
\end{equation}
where we used \eqref{eqn:similarity_Frenet}. Setting
\begin{equation}\label{eqn:Ttheta_TN}
\begin{split}
& \frac{d\overline\gamma}{d\overline{\theta}} = \overline{T}(\overline{\theta}) = P(\theta)T + Q(\theta)N,\\[2mm] 
& \frac{d\overline{\theta}}{d\theta} = 1+\epsilon\mu(\theta),
\end{split}
\end{equation}
we have from \eqref{eqn:gammabar_TN} and \eqref{eqn:Ttheta_TN}:
\begin{equation}\label{eqn:smooth_PQ}
P(\theta) = \frac{1+\epsilon \phi}{1+\epsilon\mu} = 1+\epsilon(\phi - \mu),\quad
Q(\theta) = \frac{\epsilon \psi}{1+\epsilon\mu} = \epsilon \psi,
\end{equation}
where we omitted the higher order terms in $\epsilon$. We next compute 
$d\overline{T}(\overline\theta)/d\overline\theta$ in two ways by 
using \eqref{eqn:similarity_Frenet} and \eqref{eqn:smooth_PQ}:
\begin{equation}\label{eqn:T_theta1}
 \frac{d\overline{T}(\overline\theta)}{d\overline\theta} = 
\frac{d\overline{T}(\overline\theta)}{d\theta}\,\frac{d\theta}{d\overline{\theta}}
= \frac{1}{1+\epsilon\mu}\left(\frac{dP}{d\theta} - uP-Q\right)T
+ \frac{1}{1+\epsilon\mu}\left(\frac{dQ}{d\theta} - uQ+P\right)N,
\end{equation}
where we used \eqref{eqn:similarity_Frenet}, then \eqref{eqn:smooth_PQ}.
On the other hand, using \eqref{eqn:smooth_PQ} then \eqref{eqn:similarity_Frenet}, we have
\begin{equation}\label{eqn:T_theta2}
 \frac{d\overline{T}(\overline\theta)}{d\overline\theta}
=-u\overline{T} + \overline{N} = (-\overline{u}P-Q)T + (-\overline{u}Q+P)N.
\end{equation}
Comparing \eqref{eqn:T_theta1} and \eqref{eqn:T_theta2}, we have
\begin{equation}\label{eqn:P_Q_u}
 -\overline{u}P-Q =  \frac{1}{1+\epsilon\mu}\left(\frac{dP}{d\theta} - uP-Q\right),
\quad
 -\overline{u}Q+P = \frac{1}{1+\epsilon\mu}\left(\frac{dQ}{d\theta} - uQ+P\right).
\end{equation}
In order to determine $\overline{u}$ consistently, $P$ and $Q$ must satisfy the 
equation obtained from \eqref{eqn:P_Q_u} by eliminating $\overline{u}$:
\begin{equation}
 \frac{dP}{d\theta} Q - P\frac{dQ}{d\theta} + \epsilon\mu (P^2+Q^2)=0.
\end{equation}
Substituting \eqref{eqn:smooth_PQ}, we obtain from the $O(\epsilon)$ term
\begin{equation}
 -\frac{d\psi}{d\theta} + \mu = 0.
\end{equation}
Thus, $\overline{u}$ and $\overline{\theta}$ are seen to be
\begin{equation}\label{eqn:var_u_theta}
\begin{split}
& \overline{u} = u+\epsilon\delta u,\quad 
\delta u = -\left\{\frac{d\psi}{d\theta} u + \frac{d}{d\theta}\left(\phi-\frac{d\psi}{d\theta}\right)\right\},\\[2mm]
&d\overline{\theta} = d\theta + \epsilon \delta(d\theta),\quad \delta(d\theta) = \frac{d\psi}{d\theta}\,d\theta.
\end{split}
\end{equation}
We compute the variation of $q$ by using $q^2=\langle T,T\rangle$, where $\langle\cdot,\cdot\rangle$
is the Euclidean inner product, so that $\delta(q^2)=2\langle \delta T,T\rangle$. Then, $\delta T$
is computed from \eqref{eqn:Ttheta_TN} and \eqref{eqn:T_theta1} as
\begin{displaymath}
 \delta T = 
\lim_{\epsilon \to 0}\frac{(P-1)T + QN}{\epsilon} = 
\left(\phi - \frac{d\psi}{d\theta}\right)T + \psi N.
\end{displaymath}
Therefore, we have
\begin{equation}\label{eqn:var_q}
 \frac{\delta q}{q} = \phi - \frac{d\psi}{d\theta}.
\end{equation}

Now, we are ready to calculate the variation of $\mathcal{F}^{\lambda,a}$:
\begin{equation}\label{eqn:var_F}
\delta\mathcal{F}^{\lambda,a}(\gamma)
= \int_{\theta_1}^{\theta_2}\frac{1}{2}
\left\{
2a^2 u\, \delta u + \lambda\delta\left(\left(\frac{q_1\,q_2}{q^{2}}\right)^a\right)
\right\}
\,(1+\delta\theta)\,\mathrm{d}\theta.
\end{equation}
We have, by virtue of \eqref{eqn:var_q}, 
\begin{equation}\label{eqn:var_2ndterm}
 \delta\left(\left(\frac{q_1\,q_2}{q^{2}}\right)^a\right)
= -2a \left(
 \frac{\delta q}{q} - \frac{1}{2}\frac{\delta q_1}{q_1} - \frac{1}{2}\frac{\delta q_2}{q_2}
\right)\left(\frac{q_1\,q_2}{q^{2}}\right)^a
= -2a\left(\widetilde\phi - \frac{d\widetilde\psi}{d\theta}\right),
\end{equation}
where
\begin{equation}\label{eqn:tilde_phi_psi}
\begin{split}
& \widetilde\phi = \phi - \frac{\phi_1 + \phi_2}{2},\quad
 \widetilde\psi = \psi - \frac{\psi_1 + \psi_2}{2},\\[2mm] 
&\phi_i = \phi(\theta_i),\quad \psi_i = \psi(\theta_i),\quad (i=1, 2).
\end{split}
\end{equation}
Then, we obtain the {\em first variation formula} from \eqref{eqn:var_F} after some straightforward
calculations by using \eqref{eqn:var_u_theta} and \eqref{eqn:var_2ndterm}:
\begin{equation}\label{eqn:var_F2}
 \delta\mathcal{F}^{\lambda,a}(\gamma)
= - \frac{1}{2}\left[ a^2u\left(\widetilde\phi-\frac{d\widetilde\psi}{d\theta}\right)
+ H(\gamma)\widetilde\psi\,\right]_{\theta_1}^{\theta_2} + 
\frac{a}{2}\int_{\theta_1}^{\theta_2}\left\{au'-\lambda\left(\frac{q_1q_2}{q^2}\right)^a\right\}
\left(\widetilde\phi-\frac{d\widetilde\psi}{d\theta}+u\widetilde\psi\right)\,{\rm d}\theta, 
\end{equation}
where
\begin{equation}\label{eqn:H}
 H(\gamma) = a^2 u(\theta)^2 - \lambda\left(\frac{q_1\,q_2}{q^{2}}\right)^a.
\end{equation}
The first variation formula implies that if $\gamma$ is a critical point of the fairing energy for
deformations which respect the boundary condition eliminating the first term in \eqref{eqn:var_F2},
then $\gamma$ satisfies
\begin{equation}\label{eqn:st-Burgers2}
 au'-\lambda\left(\frac{q_1q_2}{q^2}\right)^a=0,
\end{equation}
which is equivalent to the Riccati equation for qAC \eqref{eqn:Riccati_LAC} together with
\eqref{eqn:C-H}. Indeed, elimination of $\lambda$ via differentiation and evaluation modulo
\eqref{eqn:C-H} lead to the stationary Burgers equation obtained by differentiating
\eqref{eqn:Riccati_LAC}. Hence, the parameter $\lambda$ plays the role of a constant of integration.

Let us examine the boundary condition required in the above computation.  Noticing that $H(\gamma)$
is a first integral of \eqref{eqn:st-Burgers2}, we put $H(\gamma)=C = \mbox{const}$.  Then, the
boundary term in \eqref{eqn:var_F2} gives
\begin{equation}
\left[ a^2u\Big(\widetilde\phi-\frac{d\widetilde\psi}{d\theta}\Big)
+ H(\gamma)\widetilde\psi\,\right]_{\theta_1}^{\theta_2}=
 a^2(u_1+u_2)\left(\frac{\phi_1-\phi_2}{2} 
- \frac{\left.\frac{d\psi}{d\theta}\right|_{\theta=\theta_1} - 
\left.\frac{d\psi}{d\theta}\right|_{\theta=\theta_2}}{2}\right)
+C(\psi_2-\psi_1),
\end{equation}
where $u_i=u(\theta_i)$, $i=1, 2$. If we require preservation of the total turning angle, that
is, $\delta(\theta_2-\theta_1)=0$, which is the analogue of the preservation of arc length in
Euclidean geometry, then it follows from \eqref{eqn:var_u_theta} that $\psi_1=\psi_2$ so
that, by virtue of \eqref{eqn:var_q}, the boundary term vanishes if
\begin{equation}
  \frac{\delta q_1}{q_1} = \frac{\delta q_2}{q_2}.
\end{equation}
Hence, we conclude that $q_2/q_1$, namely, the ratio of length of tangent vectors at the endpoints
is preserved by the variation. Note that this condition is invariant with respect to similarity
transformations. Summarizing the discussion above, we obtain the following theorem:
%
\begin{thm}\label{mainresult}
If a plane curve $\gamma$ is a critical point of the fairing energy $\mathcal{F}^{\lambda,a}$
\eqref{eqn:fairing_functional} under the assumption of preservation of the total turning angle and
the boundary condition that the ratio of length of tangent vectors at the endpoints is preserved,
then the similarity curvature $u$ satisfies $u'=au^2+c$, where $c$ is a constant. Therefore, quasi
aesthetic curves of slope $\alpha\not=1$ are critical points of the fairing functional.
\end{thm}
%
\section{Discrete LAC and qAC}\label{sec:dLAC}
\begin{center}
\begin{figure}[t]
\begin{center}
\includegraphics[scale=0.5]{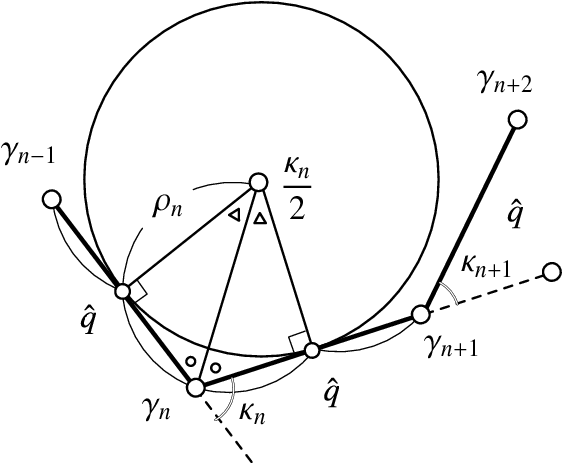}\hskip60pt
\includegraphics[scale=0.5]{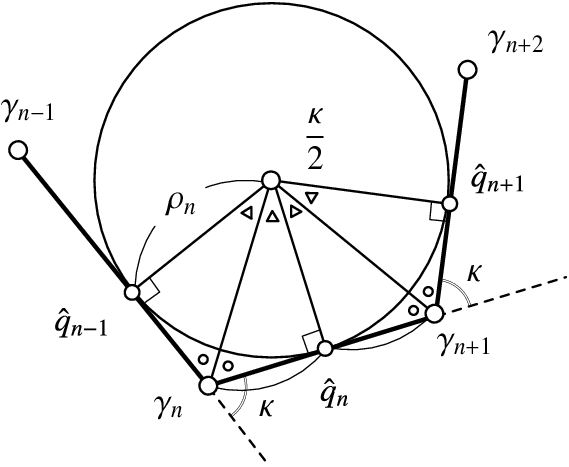}
\end{center}
\caption{Radii of osculating circles of an arc length parameterized discrete plane curve in the
Euclidean geometry and a discrete curve of constant turning angle in the similarity geometry.  Left:
Euclidean geometry, $|\gamma_{n+1}-\gamma_n|= \hat q_n=\hat q = \mbox{const}$.  Right:
similarity geometry, $ \angle(\gamma_{n}-\gamma_{n-1},
\gamma_{n+1}-\gamma_{n})=\kappa_n=\kappa = \mbox{const}$.  In both cases, the radii are given
by $\rho_n=(\hat q_n/2)\cot(\kappa_n/2)$ with $\hat q_n=q_n$ in the case of similarity
geometry. }\label{fig:radii}
\end{figure}
\end{center}

One of the benefits of the formulation developed in the preceding sections is that one is led to the
construction of a natural discrete analogue of LAC and qAC which preserves the underlying integrable
nature of these curves. It is expected that these discrete curves obtained on the principle of
structure preservation have better quality as discrete curves compared to other existing
discretizations regarded as approximations (cf.\ Section 7). In this section, we construct the
discrete analogue of LAC and qAC by using the framework of integrable evolution of discrete plane
curves in similarity geometry as discussed in \cite{KKM}.
\begin{center}
\begin{figure}[t]
\begin{center}
\includegraphics[scale=0.8]{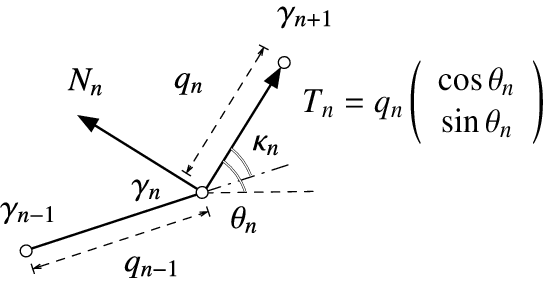}
\end{center}
\caption{Description of discrete plane curves in similarity geometry.}\label{fig:discrete_similarity_frame}
\end{figure}
\end{center}
Let $\gamma_n\in\mathbb{R}^2$, $n\in\mathbb{Z}$ be a discrete plane curve. As shown in Figure
\ref{fig:discrete_similarity_frame}, we introduce the {\em similarity Frenet frame} $F_n\in{\rm
CO}^{+}(2)$ according to
\begin{equation}\label{eqn:discrete_similarity_frame}
 F_n = (T_n,N_n),\quad  T_n = \gamma_{n+1} - \gamma_n,\quad N_n = JT_n,
\end{equation}
where $T_n$ and $N_n$ are discrete tangent and normal vectors, respectively, and we write
\begin{equation}
 q_n = |T_n| = \sqrt{\langle T_n,T_n\rangle}.
\end{equation}
Then, $F_n$ satisfies the {\em discrete similarity Frenet formula}
\begin{equation}\label{eqn:d_similarity_Frenet}
\begin{split}
& F_{n+1} = F_n L_n,\quad L_n = u_nR(\kappa_{n+1}),\quad
R(\kappa_{n+1})=\left(\begin{array}{cc}
\cos\kappa_{n+1}  &-\sin\kappa_{n+1} \\
\sin\kappa_{n+1} & \cos\kappa_{n+1}
\end{array}\right),\\[2mm]
& u_n = \frac{q_{n+1}}{q_n},\quad \kappa_n = \angle (T_{n-1},T_n),
\end{split}
\end{equation}
where $u_n$ plays the role of a discrete counterpart of the similarity curvature of smooth plane
curves. Hereafter, we assume that the discrete turning angle $\kappa_n = \kappa = \mbox{const.}$,
and the associated discrete curves may be regarded as the similarity geometric analogues of arc
length parameterized discrete curves in Euclidean geometry. Such discrete curves may
be referred to as ``{\em similarity arc length parametrized}''.
%
\begin{rem}\label{rem:discrete_curvature}
There is an interesting correspondence on the radii of osculating circles for both arc length
parameterized discrete curves in the Euclidean geometry and the similarity geometry (see Figure
\ref{fig:radii}). In the Euclidean geometry, a discrete plane curve $\gamma_n\in\mathbb{R}^2$ is
said to be an arc length parameterized discrete curve if the segment length is constant
\cite{Hoffmann:MI}, i.e., $|\gamma_{n+1}-\gamma_n|=\hat q_n = \hat q= \mbox{const}$. Then, there
exists a circle touching the two segments $\gamma_{n}-\gamma_{n-1}$ and $\gamma_{n+1}-\gamma_n$ at
their midpoints, and its radius $\rho_n$ is given by $\rho_n = ({\hat q}/2)\cot(\kappa_n/2)$, where
$\kappa_n = \angle(\gamma_{n}-\gamma_{n-1}, \gamma_{n+1}-\gamma_{n})$.  On the other hand, in the
similarity geometry, there exists a circle touching simultaneously the three consecutive segments
$\gamma_{n}-\gamma_{n-1}$, $\gamma_{n+1}-\gamma_n$, $\gamma_{n+2}-\gamma_{n+1}$ with the second
segment being touched at its midpoint. The radius of the circle $\rho_n$ is given by $\rho_n = (\hat
q_n/2) \cot(\kappa/2)$, which is the same expression as in the Euclidean case. Note that $\hat
q_n=q_n$ in the case of similarity geometry. From this observation one may regard $1/\rho_n$ as a
discrete analogue of the Eulidean curvature, and one can trace its change along the discrete curve
by $1/q_n$.
\end{rem}

We consider a discrete (time) evolution of a discrete curve $\gamma_n$ preserving the constant
turning angle $\kappa_n=\kappa$. We denote the original discrete curve by $\gamma_n^0$ and the curve
obtained after $m$ discrete time steps is labelled by $\gamma_n^m$. The quantities relevant to these
discrete curves are written in a similar manner. For example, $q_n^m = |\gamma_{n+1}^m-\gamma_n^m|$
and $u_n^m = q_{n+1}^m/q_n^m$. Then, the simplest evolution is known to be given by \cite{KKM}
\begin{equation}
 \gamma_n^{m+1} = -\gamma_n^m + \frac{\sigma}{\kappa^2}
\left\{\left(\frac{1}{u_{n-1}^m}-\cos\kappa\right)T_n^m + \sin\kappa\,N_n^m\right\},
\end{equation}
where the frame $F_n^m$ satisfies
\begin{equation}\label{eqn:discrete_deformation_frame}
\begin{split}
& F_{n+1}^m = F_n^m L_n^m, \quad L_n^m=u_n^m R(\kappa),\\[2mm]
& F_n^{m+1} = F_n^m M_n^m,\quad M_n^m = H_n^mI,\quad I:\text{identity matrix},\quad
H_n^m = 1 + \frac{\sigma}{\kappa^2}\left(u_n^m - 2\cos\kappa + \frac{1}{u_{n-1}^m}\right),
\end{split}
\end{equation}
and $\sigma$ is a constant. Note that the first equation is nothing but the discrete similarity
Frenet formula.  The compatibility condition of \eqref{eqn:discrete_deformation_frame},
$L_{n}^{m+1}M_n^m = M_{n+1}^mL_n^m$, yields the {\em discrete Burgers equation}
\cite{Hirota:Burgers,Nishinari-Takahashi}
\begin{equation}\label{eqn:dBurgers}
 \frac{u_n^{m+1}}{u_n^m}
= \frac{1 + \frac{\sigma}{\kappa^2}\left(u_{n+1}^m - 2\cos\kappa + \frac{1}{u_{n}^m}\right)}
{1 + \frac{\sigma}{\kappa^2}\left(u_{n}^m - 2\cos\kappa + \frac{1}{u_{n-1}^m}\right)},
\end{equation}
which is linearized in terms of $q_n^m$ to
\begin{equation}\label{eqn:dDiffusion}
 \frac{q_n^{m+1}-q_n^m}{\sigma} = \frac{q_{n+1}^m -2\cos\kappa\,q_n^m +q_{n-1}^m }{\kappa^2}.
\end{equation}
Note that the continuum limit \eqref{eqn:Burgers} of \eqref{eqn:dBurgers} with $b=0$ 
is obtained by setting
\begin{equation}\label{eqn:cont_limit_u}
\begin{split}
& u_n^m = 1 - \kappa u,\quad \theta = n\kappa,\quad \kappa\to 0,\\
& t = m\sigma,\quad \sigma \to 0.
\end{split}
\end{equation}
Imposing the stationarity ansatz $u_n^{m+1}=u_n^m$ on the discrete Burgers equation 
and neglecting the superscript $m$, we obtain the discrete stationary Burgers equation
\begin{equation}\label{eqn:d_stationaryBurgers}
u_{n+1}  + \frac{1}{u_{n}}
= 
u_{n} + \frac{1}{u_{n-1}},
\end{equation}
whose continuum limit gives the stationary Burgers equation
\begin{equation}
 \frac{d^2u}{d\theta^2} = 2u\frac{du}{d\theta}.
\end{equation}
Equation \eqref{eqn:d_stationaryBurgers} can be integrated to yield the discrete
Riccati equation
\begin{equation}\label{eqn:dRiccati}
 u_{n+1}  + \frac{1}{u_{n}} = C,
\end{equation}
where $C$ is an integration constant. The existence of the continuum limit of \eqref{eqn:dRiccati}
requires the parametrization $C=2-c\kappa^2$, leading to
\begin{equation}
 \frac{du}{d\theta} = u^2 + c.
\end{equation}
In order to construct the discrete analogue of \eqref{LAC} and \eqref{eqn:Riccati_LAC}, 
we replace $u_n$ by $(u_n)^a$, where $a=\alpha-1$, to obtain
\begin{equation}\label{eqn:d_stationary_Burgers_a}
 (u_{n+1})^a  + \frac{1}{(u_{n})^a}
= 
(u_{n})^a + \frac{1}{(u_{n-1})^a},
\end{equation}
and
\begin{equation}\label{eqn:dRiccati_a}
  (u_{n+1})^a  + \frac{1}{(u_{n})^a} = C,
\end{equation}
respectively. This $a$ dependence is consistent with the parametrization \eqref{eqn:cont_limit_u}
which comes from a geometric restriction on the continuum limit. Actually, noticing that $(u_n)^a =
(1-\kappa u)^a = 1-au\kappa + O(\kappa^2)$, we see that \eqref{eqn:d_stationary_Burgers_a} and
\eqref{eqn:dRiccati_a} reduce to \eqref{LAC} and \eqref{eqn:Riccati_LAC}, respectively, if we set
$C=2-ac\kappa^2$.  Let us consider the solution of \eqref{eqn:dRiccati_a}, which may be linearized
according to
\begin{equation}\label{eqn:dlinear_p}
 \frac{p_{n+1} -2 p_{n} + p_{n-1}}{\kappa^2} = -ac p_{n}
\end{equation}
by putting
\begin{equation}
 u_n = \left(\frac{p_{n+1}}{p_n}\right)^{\frac{1}{a}}.
\end{equation}
In the case $c=0$, the solution of \eqref{eqn:dlinear_p} is given by $p_n = c_1n + c_2$ with $c_1,c_2$ being arbitrary constants to yield
\begin{equation}\label{eqn:dLAC}
 u_n = \left(1+\frac{a\lambda\kappa}{a\lambda\kappa n+1}\right)^{\frac{1}{a}},
\end{equation}
where $\lambda = c_1/(\kappa ac_2)$. It is evident that \eqref{eqn:dLAC} yields the original
expression for the similarity curvature of LAC \eqref{eqn:LAC_similarity_curvature} by applying the
continuum limit \eqref{eqn:cont_limit_u}. The above discussion motivates the following natural
definition.
%
\begin{defn}\label{defn:dLAC}
Let $\gamma_n$ be a discrete plane curve of constant turning angle $\kappa$.
$\gamma_n$ is said to
be a discrete LAC (dLAC) of slope $\alpha$ if $u_n$ satisfies
\begin{equation}
  (u_{n+1})^a  + \frac{1}{(u_{n})^a} = 2.
\end{equation}
$\gamma_n$ is said to be a discrete qAC (dqAC) of slope $\alpha$ if
$u_n$ satisfies
\begin{equation}
  (u_{n+1})^a  + \frac{1}{(u_{n})^a} = C,\quad C\in\mathbb{R}.
\end{equation}
In both cases, $a = \alpha - 1$. 
\end{defn}

\begin{figure}[t]
\begin{center}
\includegraphics[width=6.5cm]{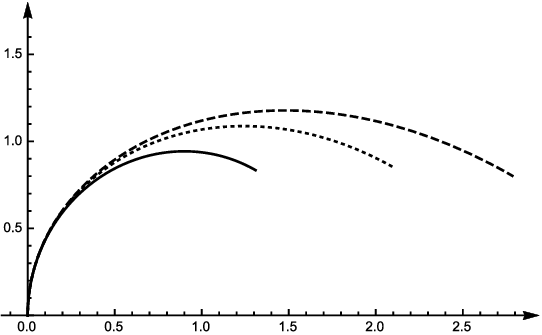}
\qquad \includegraphics[width=6.5cm]{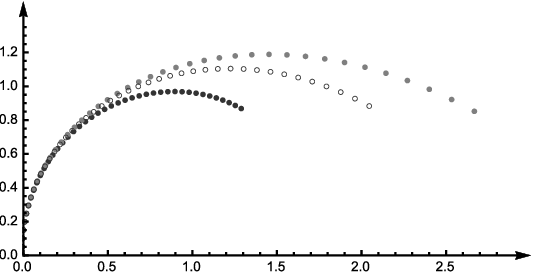}
\end{center}
\caption{Smooth and discrete qAC. Left: qAC with (i) $(a,c) = (1,0)$ (solid line), (ii) $(3/2,-1)$
(dashed line), (iii)$(3,-2/3)$ (dotted line).  Right: dqAC with the same parameters (i): black,
(ii): gray (iii): white, where $C = 2 - ac\kappa^2$ and $\kappa=0.05$.}\label{fig:qAC}
\end{figure}

Figure \ref{fig:qAC} illustrates some qAC and dqAC with the same parameters $a$ and $c$. 
%
\section{Variational formulation of dLAC and dqAC}
It turns out that, as in the continuous case, the dLAC and the dqAC proposed in Section
\ref{sec:dLAC} may be obtained via a variational principle. Indeed, in the following, we demonstrate
that the dLAC and the dqAC may be characterized as the stationary curves of constant turning angle
of the {\em discrete fairing energy functional} $\varPhi^{\lambda,a}$ given by
\begin{equation}\label{eqn:discrete_fairing_functional}
 \varPhi^{\lambda,a}(\gamma) = \sum_{n=n_1}^{n_2-1}
\left\{(u_n)^a + \frac{1}{(u_n)^a} + \lambda\left(\frac{q_{n_1}q_{n_2}}{q_nq_{n+1}}\right)^a\right\},
\end{equation}
with respect to an arbitrary variation of the discrete curve $\gamma_n$ which we write as
\begin{equation}\label{eqn:variation_dgamma}
\delta\gamma_n=\xi_nT_n + \eta_nN_n.
\end{equation}
To this end, we first compute the variation of the frame by using the discrete similarity Frenet
formula \eqref{eqn:d_similarity_Frenet} as
\begin{equation}
\delta T_n = \delta\gamma_{n+1} - \delta\gamma_n = \chi_n T_n + \psi_n N_n,\quad
\delta N_n =  -\psi_n T_n + \chi_n N_n,
\end{equation}
or
\begin{equation}\label{eqn:var_dFrame}
\delta F_n = F_n M_n,\quad 
M_n = \left(\begin{array}{cc}\chi_n & \psi_n \\ -\psi_n  &\chi_n \end{array}\right),
\end{equation}
where
\begin{equation}
\begin{split}
& \chi_n = \xi_{n+1} u_n\cos\kappa_{n+1}-\eta_{n+1}u_n\sin\kappa_{n+1}-\xi_n,\\
& \psi_n = \xi_{n+1} u_n\sin\kappa_{n+1}+\eta_{n+1}u_n\cos\kappa_{n+1}-\eta_n.
\end{split}
\end{equation}
The variation of the frame \eqref{eqn:var_dFrame} must be compatible with the 
similarity Frenet formula \eqref{eqn:d_similarity_Frenet}. Accordingly, the associated
 compatibility condition
$\delta L_n = L_nM_{n+1} - M_nL_n$ results in the pair
\begin{equation}
 \begin{split}
  & \frac{\delta u_n}{u_n}\cos\kappa_{n+1} -\delta\kappa_{n+1}\sin\kappa_{n+1}
= \cos\kappa_{n+1}(\chi_{n+1}-\chi_n) - \sin\kappa_{n+1}(\psi_{n+1}-\psi_n),\\[2mm]
& \frac{\delta u_n}{u_n}\sin\kappa_{n+1} +\delta\kappa_{n+1}\cos\kappa_{n+1}
= \cos\kappa_{n+1}(\psi_{n+1}-\psi_n) + \sin\kappa_{n+1}(\chi_{n+1}-\chi_n),
 \end{split}
\end{equation}
from which we obtain the variation of $u_n$ and $\kappa_n$ as
\begin{equation}\label{eqn:var_un}
 \frac{\delta u_n}{u_n} = \chi_{n+1} - \chi_n,\quad \delta\kappa_{n+1} = \psi_{n+1}-\psi_n.
\end{equation}
Taking the variation of $q_n^2 = \langle T_n,T_n\rangle$, we have $2q_n\delta q_n=2\langle \delta
T_n,T_n\rangle$. Then, from \eqref{eqn:var_dFrame}, we obtain the variation of $q_n$ as
\begin{equation}\label{eqn:var_qn}
 \frac{\delta q_n}{q_n} = \chi_n.
\end{equation}
Note that $\delta u_n$ can also be calculated by using $u_n = q_{n+1}/q_n$, which is consistent
with \eqref{eqn:var_un}.

On use of the variations \eqref{eqn:var_un} and \eqref{eqn:var_qn}, the variation of the discrete
fairing energy functional is seen to be
\begin{align}
\delta\varPhi^{\lambda,a}(\gamma) 
&= a\sum_{n=n_1}^{n_2-1}
\left[\left\{(u_n)^a - \frac{1}{(u_n)^a}\right\}\frac{\delta u_n}{u_n}  
+ \lambda\left(
-\frac{\delta q_n}{q_n} - \frac{\delta q_{n+1}}{q_{n+1}} 
+ \frac{\delta q_{n_1}}{q_{n_1}} + \frac{\delta q_{n_2}}{q_{n_2}}
\right)\left(\frac{q_{n_1}q_{n_2}}{q_nq_{n+1}}\right)^a\right]\nonumber\\
&= a\sum_{n=n_1}^{n_2-1}
\left[\left\{(u_n)^a - \frac{1}{(u_n)^a}\right\}(\widetilde\chi_{n+1} - \widetilde\chi_n)
- \lambda \left(\frac{q_{n_1}q_{n_2}}{q_nq_{n+1}}\right)^a (\widetilde\chi_{n+1}
+ \widetilde\chi_n)\right]\nonumber\\
&=-a\sum_{n=n_1+1}^{n_2-2} \left\{1+\frac{1}{(u_{n-1}u_{n})^a}\right\}
\left\{(u_n)^a - (u_{n-1})^a +\lambda \left(\frac{q_{n_1}q_{n_2}}{q_{n-1}q_{n}}\right)^a\right\}\,
\widetilde\chi_n \nonumber\\
&+ a\left[
(u_{n_2-1})^a  - \frac{1}{(u_{n_2-1})^a}
+ (u_{n_1})^a  - \frac{1}{(u_{n_1})^a}
-\lambda\left(\frac{q_{n_1}}{q_{n_2-1}}\right)^a
+\lambda\left(\frac{q_{n_2}}{q_{n_1+1}}\right)^a
\right]\,\frac{\chi_{n_2}-\chi_{n_1}}{2}, \label{eqn:1stvar_dfairing}
\end{align}
where
\begin{equation}
\widetilde\chi_n = \chi_n - \frac{\chi_{n_1}+\chi_{n_2}}{2}.
\end{equation}
The first variation formula \eqref{eqn:1stvar_dfairing} implies that if $\gamma_n$ is a critical
point of the discrete fairing energy for deformations which respect the boundary condition, 
then $\gamma_n$ satisfies
\begin{equation}
 (u_n)^a - (u_{n-1})^a +\lambda \left(\frac{q_{n_1}q_{n_2}}{q_{n-1}q_{n}}\right)^a = 0,\quad n=n_1+1,\ldots,n_2-1,
\end{equation}
which is equivalent to \eqref{eqn:d_stationary_Burgers_a} or \eqref{eqn:dRiccati} together with
$u_n=q_{n+1}/q_n$ in the same manner as in the continuous case. The boundary term vanishes
iff $\chi_{n_1}=\chi_{n_2}$, which implies that $\delta (q_{n_1}/q_{n_2})=0$ from
\eqref{eqn:var_qn}. This means that the ratio of length of segments at the endpoints is preserved by
the variation, which is the discrete analogue of the boundary condition in the smooth curve case.
%
\begin{thm}\label{mainresult_discrete}
If a discrete plane curve $\gamma_n$ is a critical point of the discrete fairing energy
$\varPhi^{\lambda,a}$ \eqref{eqn:discrete_fairing_functional} under the boundary condition that the
ratio of length of segments at the endpoints is preserved, then $u_n$ satisfies
\eqref{eqn:d_stationaryBurgers}. Therefore, discrete quasi aesthetic curves of slope $\alpha\not=1$
are those discrete curves of constant turning angle which constitute critical points of the discrete
fairing functional.
\end{thm}
%
\begin{rem}
Since $\psi_n$ does not enter the variation \eqref{eqn:1stvar_dfairing} of the discrete fairing
functional, whether the variation of the curve preserves the constancy of the turning angle or not
does not affect the discrete Euler-Lagrange equation. However, if $\kappa_{n+1}=\kappa_n$ is to be
preserved by the variation then, by virtue of \eqref{eqn:var_un}, $\psi_n$ is no longer arbitrary
but constrained by $\psi_{n+1}-\psi_n = \mbox{const}$. It is also observed that the structure of the
variation \eqref{eqn:1stvar_dfairing} may be interpreted in a simple geometric manner. Since, up to
Euclidean motions, a discrete curve is uniquely determined by the angles $\kappa_n$ and the lengths
$q_n$ of the segments, we may regard $\delta q_n$ as independent quantities in the variation of the
energy functional. More precisely, in order to respect invariance under similarity transformations,
appropriate independent variations are given by $\delta \tilde{q}_n$, where $\tilde{q}_n =
q_n/\sqrt{q_{n_1}q_{n_2}}$. Hence, since the energy functional depends on $\tilde{q}_n$ only with
$u_n = \tilde{q}_{n+1}/\tilde{q}_n$, its variation may be expressed entirely in terms of
$\widetilde{\phi}_n=\delta\tilde{q}_n/\tilde{q}_n$. In this manner, one retrieves the variation
\eqref{eqn:1stvar_dfairing} if one takes into account that, for instance, $q_{n_1}/q_{n_2-1} =
1/\tilde{q}_{n_2-1}\tilde{q}_{n_2}$.
\end{rem}
%
\section{Generation of dLAC}
In this section, we consider the problem of ${\rm G}^1$ Hermite interpolation by using dLAC, namely,
we generate the dLAC with specified endpoints and the direction of segments (tangent vectors) at the
endpoints. This problem was formulated and solved for LAC in \cite{Yoshida-Saito:G1}. In Section
\ref{subsec:dLAC1} we present a method to generate dLAC based on the similaity geometry.  In this
formulation, we assume that the discrete curves are similarity arc length parametrized; it has a
constant turning angle, or, each angle between the adjacent segments are the same, and the segment
length $q_n$ are the variables. This implies that this method can generate dLACs without inflection,
namely ``$C$-shaped'' curves only. On the other hand, the curve segments with an inflection point,
namely ``$S$-shaped'' curves are also important in the industrial design \cite{Miura2013}.  A method
of generating LAC with an inflection point has been proposed in \cite{Miura2013} when the slope
$\alpha$ is negative.  In Section \ref{subsec:dLAC2}, we present a method to generate an $S$-shaped
dLAC based on the similarity geometry.
%
\subsection{dLAC without inflection}\label{subsec:dLAC1}
\begin{center}
\begin{figure}[t]
\begin{center}
\includegraphics[scale=0.6,viewport=70 120 570 330,clip]{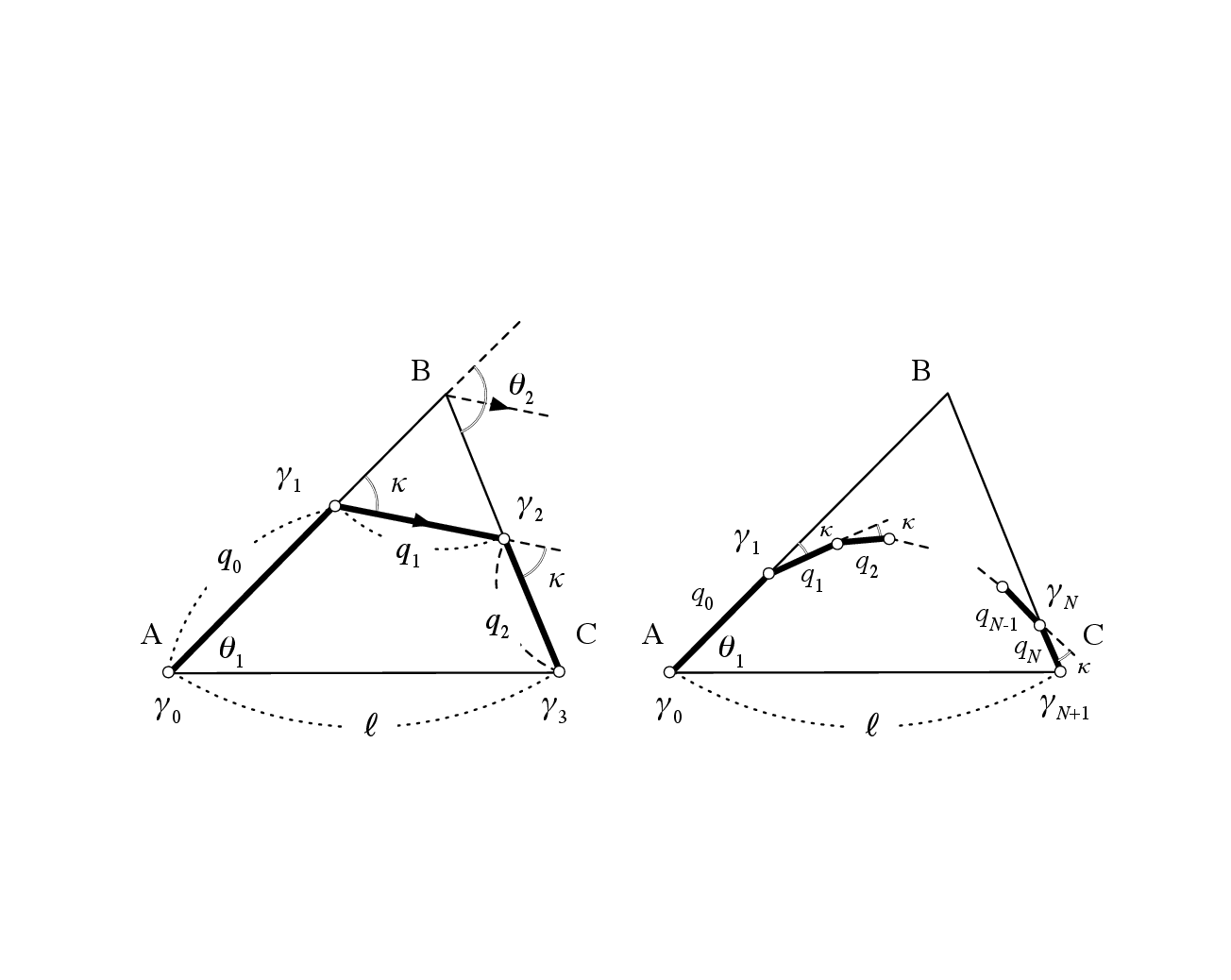}
\end{center}
\caption{Generation of dLAC by ${\rm G}^1$ interpolation. Left: four points. Right: $N+2$ points.}\label{fig:G1}
\end{figure}
\end{center}

We consider a generation method of dLAC without inflection based on the similarity geometry.  As
mentioned above, we assume that the discrete curve is similarity arc length parametrized. For
simplicity, we first construct dLAC consisting of four points for given endpoints, $\gamma_{0}$ and
$\gamma_{3}$, and the direction of the segments at those points with the specified parameter $a$.
Consider the triangle on the plane shown in Figure \ref{fig:G1}.  The problem is equivalent to
determining $\gamma_{1}$ on AB and $\gamma_2$ on BC such that $\angle (\gamma_2-\gamma_1,{\rm AB}) =
\angle ({\rm BC},\gamma_3-\gamma_2) =\kappa $, where $\kappa = \frac{1}{2} \theta_2$.  In other
words, the length of the segments $q_n=\left|\gamma_{n+1}-\gamma_n\right|$ $(n=0,1,2)$ is subject to
the constraints
\begin{align}
& q_0\cos \theta_1 + q_1\cos(\theta_1 - \kappa) + q_2\cos(\theta_1-2\kappa) = \ell, \label{eqn:G1_4pt_q2}\\
& q_0\sin \theta_1 + q_1\sin(\theta_1 - \kappa) + q_2\sin(\theta_1-2\kappa) = 0, \label{eqn:G1_4pt_q3}
\end{align}
where we have chosen the coordinates such that $\gamma_0={}^t(0,0)$ and $\gamma_3={}^t(\ell,0)$ ($\ell>0$)
without loss of generality.  Moreover, $q_n$ $(n=0,1,2)$ satisfies
\begin{equation}\label{eqn:G1_4pt_q}
(q_{0})^a - 2(q_{1})^a + (q_{2})^a = 0,
\end{equation}
for specified real number $a$. 
Therefore, the three unknown variables $q_0$, $q_1$ and $q_2$ are determined from equations
\eqref{eqn:G1_4pt_q2}, \eqref{eqn:G1_4pt_q3} and \eqref{eqn:G1_4pt_q}, in principle, and $\gamma_1$,
$\gamma_2$ are given by
\begin{equation}
\gamma_1 = \gamma_0 + q_0\left(\begin{array}{c}\cos(\theta_1-\kappa)\\\sin(\theta_1-\kappa)\end{array}\right), \quad
\gamma_2 = \gamma_1 + q_1\left(\begin{array}{c}\cos(\theta_1-2\kappa)\\\sin(\theta_1-2\kappa)\end{array}\right).
\end{equation}
It is straightforward to generalize the above procedure to generate dLAC with $N+2$ points,
$\gamma_0={}^t(0,0),\gamma_1,\ldots, \gamma_{N},\gamma_{N+1}={}^t(\ell,0)$, for given
$\gamma_0,\gamma_1$ and $\gamma_{N},\gamma_{N+1}$ being on the respective edges of the specified
triangle depicted in the second picture of Figure \ref{fig:G1}.  Then, $q_n$ ($n=0,\ldots,N$)
satisfy the following equations:
\begin{align}
&(q_{n-1})^a - 2(q_{n})^a + (q_{n+1})^a = 0,\quad n=1,\ldots, N-1,\label{eqn:G1_Npt_q}\\
&q_0\cos \theta_1 + q_1\cos(\theta_1 - \kappa) +\cdots+ q_{N}\cos(\theta_1-N\kappa)=\ell, \label{eqn:G1_Npt_q2}\\
&q_0\sin \theta_1 + q_1\sin(\theta_1 - \kappa) +\cdots+ q_{N}\sin(\theta_1-N\kappa)=0,\label{eqn:G1_Npt_q3}
\end{align}
where $\kappa=\theta_2/N$. It is possible to determine $q_n$ in principle, since we have $N+1$
equations for $N+1$ unknown variables $q_n$ ($n=0,\ldots,N$).  Then, we have
\begin{equation}
 \gamma_n = \gamma_{n-1} + q_{n-1}\left(\begin{array}{c}\cos(\theta_1-n\kappa)\\\sin(\theta_1-n\kappa)
\end{array}\right),\quad n = 1,\ldots, N. \label{eq:sum0}
\end{equation}

\begin{center}
\begin{figure}[h]
\begin{center}
\includegraphics[width=1.0\linewidth]{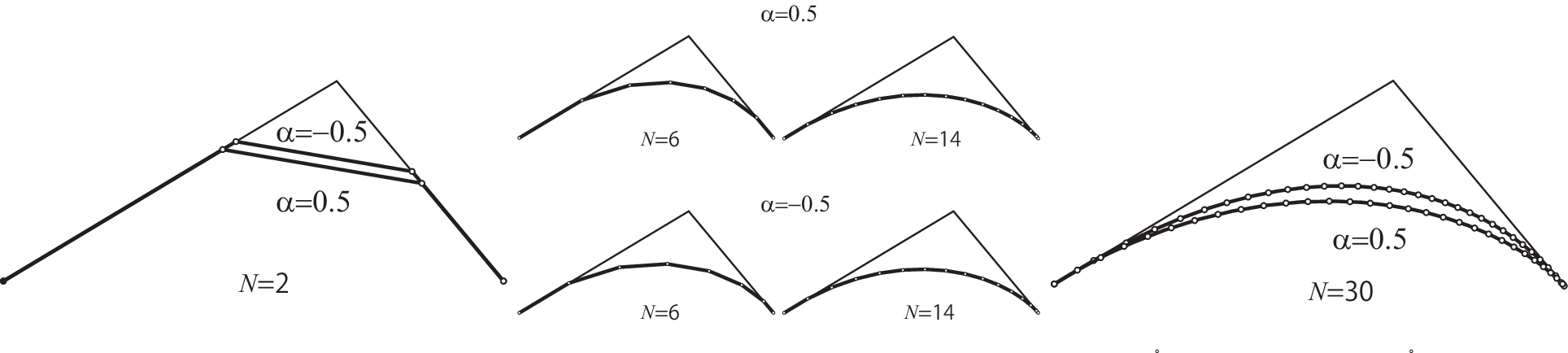} 
\end{center}
\caption{dLAC examples with $N=2,6,14,30$ for $\alpha=\pm 0.5$. }\label{fig:G1Miura}
\end{figure}
\end{center}

Now, equations \eqref{eqn:G1_Npt_q}--\eqref{eqn:G1_Npt_q3} may be solved numerically as follows:
\begin{enumerate}
\item
We may write the general solution of \eqref{eqn:G1_Npt_q} as
\begin{equation}\label{eqn:G1_Npt_qlinear}
(q_n)^a = \frac{(N-n)(q_0)^a + n(q_{N})^a}{N}\quad (n=0,\ldots,N).
\end{equation}
We also put $q_N=\beta q_0$.
\item
Substituting the above expressions into \eqref{eqn:G1_Npt_q3}, we have an equation in $\beta$.
We then solve the equation to obtain $\beta$.
\item 
Compute $q_n$ ($n=1,\ldots, N-1$) by using \eqref{eqn:G1_Npt_qlinear} to get linear expressions in terms of $q_0$.
\item 
Solve \eqref{eqn:G1_Npt_q2} for $q_0$.
\end{enumerate}
Figure \ref{fig:G1Miura} illustrates the examples of dLACs generated by the above method.  Despite
the different $\alpha$ values, the shape of the curves in the top and bottom rows of the middle
picture are similar.  When $N=2$ (total number of vertices is $4$), the triangle cut by the vertices
polyline of $\alpha=0.5$ is a little bit larger than that of $\alpha=-0.5$ as shown in the
superimposed figure on the left.  The right figure illustrates the case of $N=30$ for $\alpha=\pm
0.5$ and the area bounded by the control polyline with the curve for $\alpha=0.5$ is larger than
that with the one for $\alpha=-0.5$, that is consistent with the left and middle figures.  Each
curve reasonably approximates its continuous counterpart and the difference of those curves is
reasonable when compared with the case of continuous LAC as in \cite{Yoshida-Saito:G1}.
The discrete curvature of the curves (see Remark \ref{rem:discrete_curvature}) is
monotonically increasing from left to right and reproducing continuous LAC's property very well.
The computation time to generate dLAC on a Core i7 6700 3.4GHz is from $10$ to $20$ msec according
to $N=50$ to $300$ implemented in Matlab\textsuperscript{\textregistered}.  The computation time
based on numerical discretization of continuous LAC described in \cite{Yoshida-Saito:G1} takes about
$80$ msec in Matlab\textsuperscript{\textregistered} and the discrete implementation is much faster
since fine numerical integration to obtain the shape of the curves is not required and only coarse
summation expressed in \eqref{eq:sum0} to keep the boundary conditions is necessary.  

The above advantage may be understood to be due to the geometric characterization of dLAC themselves
as discrete curves. Namely, in the similarity geometry, the turning angles of the discrete plane
curves are constant $\kappa$, so that the shape is controlled by the segment length $q_n$. Since a
discrete analogue of the curvature is given by $(2/q_n)\tan (\kappa/2)$, which is the reciprocal of the
radius of the osculating circle touching the three consecutive edges (see Remark
\ref{rem:discrete_curvature}), if the curvature is large (resp. small) the segment length is small
(resp. large). Therefore the distribution of the vertices is dense (resp. coarse) where the
curvature is large (resp. small), which implies that the discrete plane curve under the similarity
geometry is regarded as a self-adaptive discretization. Even a coarse discrete curve can
generate sufficiently good shape. Especially, during the design stage, a designer tries to generate
as various as possible curves as to pursue the desired shape.  Coarse discrete curves are good
enough and desirable because one can generate curves quickly and check their suitability.
 
\subsection{dLAC with an inflection}\label{subsec:dLAC2}
In this section, we propose a method of generating dLAC with an inflection,
i.e. $S$-shaped dLAC based on the similarity geometry.

Unlike the case where there is no inflection as discussed in Section \ref{subsec:dLAC1},
here, uniqueness of the solution is not guaranteed.  As in the previous section, we assume
that the discrete curve has $(N+2)$-vertices
$\gamma_0={}^t(0,0),\gamma_1,\ldots,\gamma_N,\gamma_{N+1}={}^t(\ell,0)$ ($\ell>0$). Suppose
that for given $n\in\{1,2,\ldots,N-1\}$ the turning angles at the vertices
$\gamma_1,\ldots,\gamma_n$ are a constant $-\kappa\ (<0)$, and those at $\gamma_{n+1},\ldots,\gamma_N$
are $\kappa\ (>0)$.  The edge $\gamma_n - \gamma_{n+1}$ corresponds to the ``{\em inflection edge}''
where the turning angles change the sign at the left and right vertices. We put $N-n=m$ so that
there are $n+1$ vertices to the left of the inflection edge and $m+1$ vertices on the right.
\begin{figure}[h]
\begin{center}
 \includegraphics[width=0.7\textwidth]{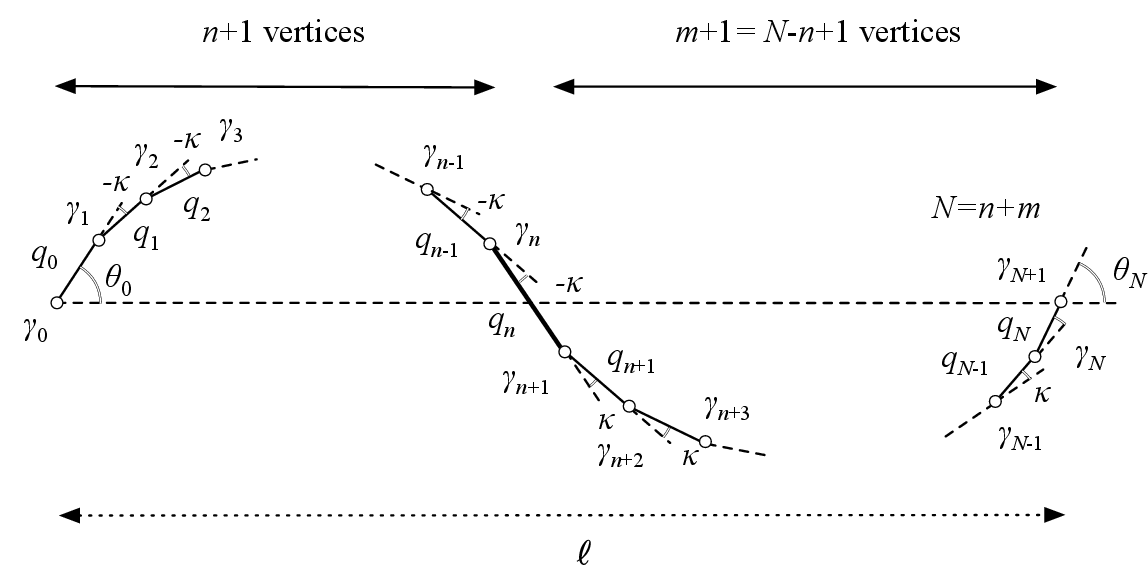}
\caption{dLAC with an inflection.} 
\end{center}
\end{figure}
The Euclidean curvature $\kappa$ of smooth LAC with an inflection point is
given by \cite{Miura2013}
\begin{equation}
 \kappa(s) = 
\left\{
\begin{array}{ll}
 (c_0s+c_1)^{-\frac{1}{\alpha}}& c_0s+c_1\geq 0,\\
-(-c_0s-c_1)^{-\frac{1}{\alpha}}& \text{otherwise},
\end{array}\right.
\end{equation}
where $c_0$, $c_1$ are parameters. For the similarity curvature of the LAC
\eqref{eqn:LAC_similarity_curvature}, the unsigned curvature radius is computed by using
\eqref{eqn:C-H} as
\begin{equation}
 |q| = \left\{
\begin{array}{ll}
z(-\hat\theta+\theta_I)^{\frac{1}{a}}&  \hat\theta\leq\theta_I,\\
z(\hat\theta-\theta_I)^{\frac{1}{a}} & \hat\theta >\theta_I,
\end{array}\right.
\end{equation}
where
\begin{equation}
 \theta = \left\{
\begin{array}{cl}
 2\theta_I - \hat\theta& \hat\theta\leq \theta_I,\\
 \hat\theta& \hat\theta> \theta_I,
\end{array}
\right.
\end{equation}
and $a=\alpha-1$, $z>0$, $\theta_I$ are parameters. Moreover, $\theta_I$ corresponds to the value of the
angle function at the inflection point. As mentioned in Remark \ref{rem:discrete_curvature}, a
discrete analogue of the curvature radius for the similarity arc length parametrized discrete plane
curve is given by $(q_k/2)\cot (\kappa/2)$, where $q_k$ is the segment length. In view of this and
Definition \ref{defn:dLAC} applied to the two parts of the dLAC, we introduce $q_k$
($k=0,1,\ldots,N$) as
\begin{equation}\label{eqn:dLAC_q}
q_k =
\begin{cases}
z(n - k + \delta)^{\frac{1}{a}}, & k = 0, \ldots, n-1, \\
\qquad z\delta^{\frac{1}{a}} & k=n,\\
z(-n + k + \delta)^{\frac{1}{a}}, & k = n+1, \ldots, N,
\end{cases}
\end{equation}
where $z,\delta >0 $ are parameters to be determined. Then $q_k$ ($k=0,\ldots,N$) satisfies the
following equations:
\begin{align}
& \sum_{i=0}^{n-1} q_i \cos (\theta_0 - i \kappa) + q_{n}\cos(\theta_0-n\kappa) 
+ \sum_{j=1}^{m} q_{n + j} \cos (\theta_0 - n \kappa +  j \kappa) = \ell, \label{eq:inflection_horizontal} \\
& \sum_{i=0}^{n-1} q_i \sin (\theta_0 - i \kappa) + q_{n}\sin(\theta_0-n\kappa) 
+ \sum_{j=1}^{m} q_{n + j} \sin (\theta_0 - n \kappa +  j \kappa) = 0, \label{eq:inflection_vertical} \\
& \theta_0 - n \kappa + m \kappa = \theta_N \label{eqn:bdry_angle_equation}. 
\end{align}
From \eqref{eqn:bdry_angle_equation} we have
\begin{equation}\label{eqn:angle_condition}
\kappa = \frac{\theta_N -  \theta_0}{m - n }.
\end{equation}
For a given number of vertices $N+2$, the slope $a = \alpha - 1$, the endpoints $\gamma_0$,
$\gamma_{N+1}$, the angles $\theta_0, \theta_N$ at $\gamma_0, \gamma_{N+1}$, respectively, and the
index $n$ of the inflection, one can compute the pair $(\delta,z)$ by solving
\eqref{eq:inflection_horizontal} and \eqref{eq:inflection_vertical} and the dLAC with an inflection
can be generated accordingly. It should be remarked that in the discrete case, $n$ must be
prescribed and cannot be determined from the equations, so that the dLAC cannot be uniquely
determined, while in the smooth case LAC with an inflection point can be uniquely determined
under a certain moderate condition \cite{Miura2013}.

\begin{figure}[h]
 \includegraphics[bb=0 0 894 447,width=0.45\textwidth]{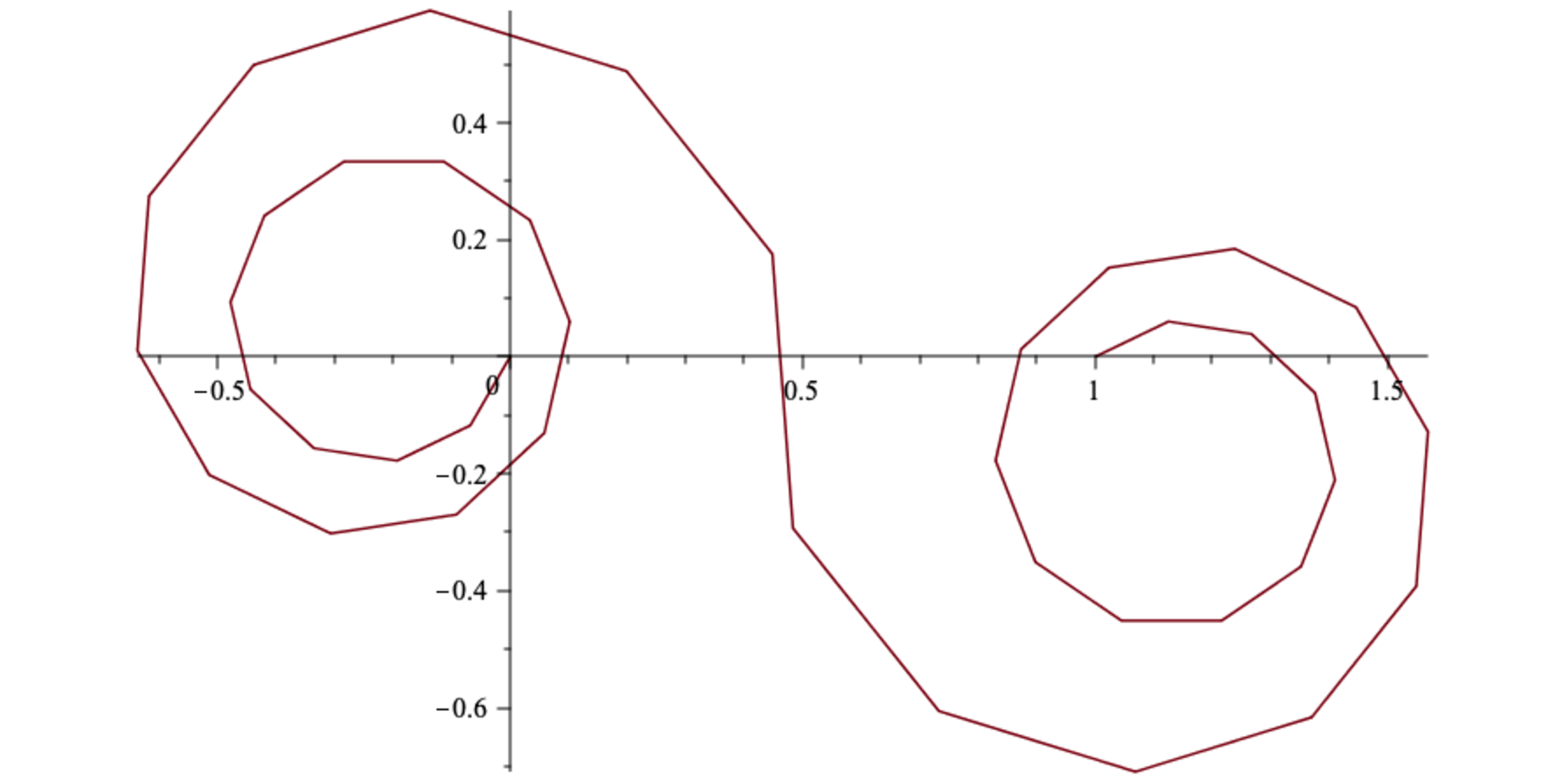}\quad
 \includegraphics[bb=0 0 894 447,width=0.45\textwidth]{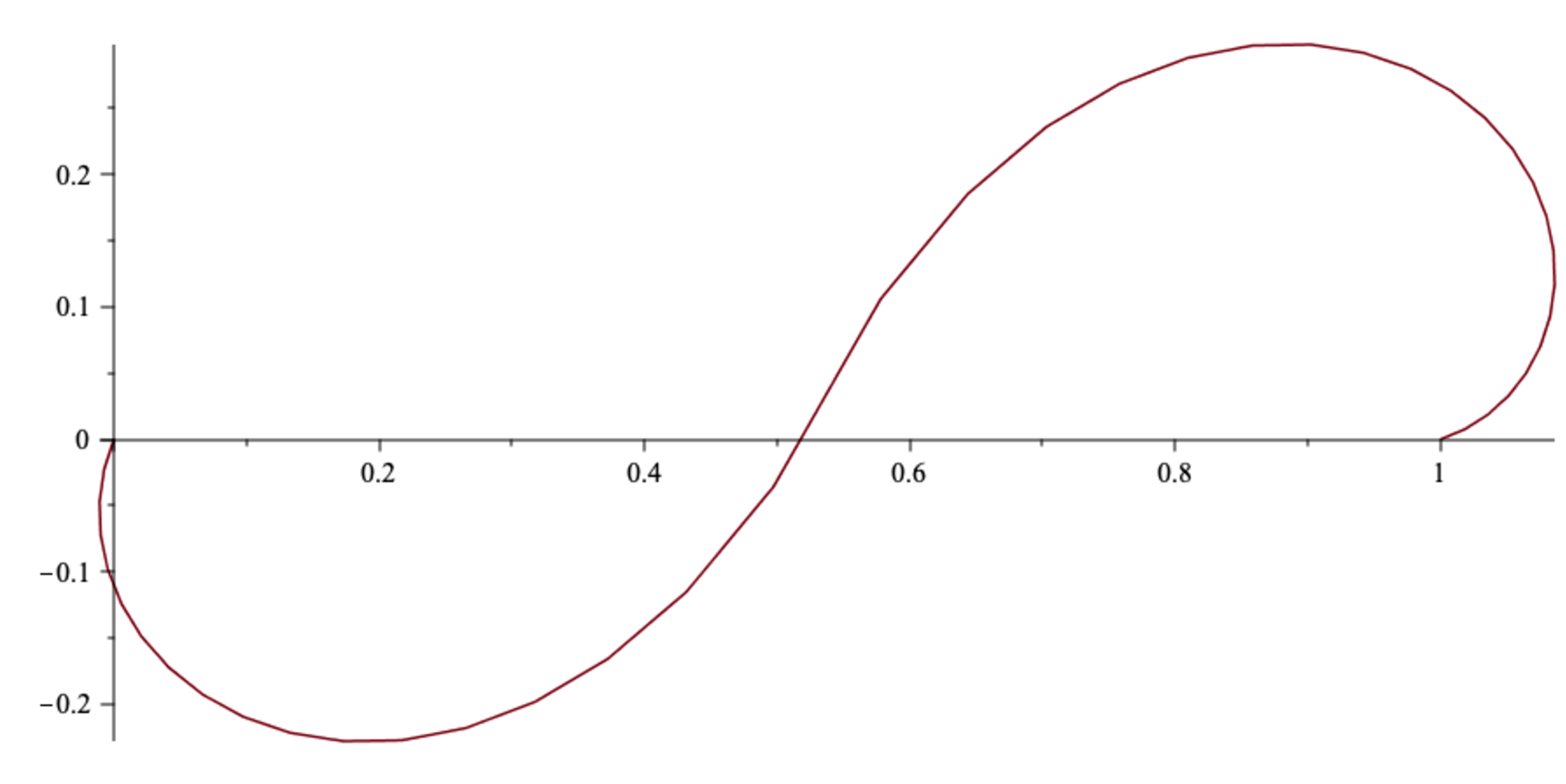}
\caption{Example of dLACs with an inflection to be excluded.}\label{fig:illegal_dLAC}
\end{figure}

However, if we prescribe the index $n$ of the inflection, sometimes there is no solution
$\delta$, or the triplet $(n,\delta,z)$ generates a discrete curve with undesirable shape as
illustrated in Figure \ref{fig:illegal_dLAC}.  Therefore, we impose the following assumptions in
order to guarantee the existence of the solution and to exclude discrete curves of the type
displayed in Figure \ref{fig:illegal_dLAC}:
\begin{align}
\label{eqn:bdry_angle_assumption1}
\theta_0 - (n - l) \kappa &\geq - \frac{\pi}{2}, \quad {\rm for \ all} \quad l = 0,1, 2, \ldots, \\
\label{eqn:bdry_angle_assumption2}
\theta_0 - n \kappa &\leq 0.
\end{align}
Equations \eqref{eqn:bdry_angle_assumption1} and \eqref{eqn:bdry_angle_assumption2} are for
excluding the dLACs on the left and on the right of Figure \ref{fig:illegal_dLAC},
respectively.  We remark, however, that it is possible to control the number of loops if
desired.  Then, we have the following restriction for the index $n$ of the
inflection, that is, even though $n$ is still not unique, it is restricted considerably in
the following manner:
\begin{lemma}
Assume that $\kappa > 0$.  Then we have
the following estimate:
\begin{equation}\label{eqn:position_estimate}
\begin{split}
& \dfrac{\theta_0 + \frac{\pi}{2}}{\theta_0 + \theta_N + \pi} N \leq 
n \leq \frac{\theta_0}{\theta_0 + \theta_N} N,\quad (\theta_0 > \theta_N),\\[2mm]
&  \frac{\theta_0}{\theta_0 + \theta_N} N\leq n \leq 
\dfrac{\theta_0 + \frac{\pi}{2}}{\theta_0 + \theta_N + \pi} N,\quad (\theta_0 < \theta_N).
\end{split}
\end{equation}
\end{lemma}
\begin{proof}
We consider the first case. From the conditions
\eqref{eqn:angle_condition}, \eqref{eqn:bdry_angle_assumption1} and $N =
n + m$, we have
\begin{align*}
0 &\leq \theta_0 - (n - l) \kappa + \frac{\pi}{2}
= \theta_0 - (n - l) \frac{\theta_N - \theta_0}{m - n} + \frac{\pi}{2} \\
&= \frac{1}{m - n} \left( -(\theta_0 + \theta_N + \pi) n 
+ (\theta_N-\theta_0)l  + \left(\theta_0 + \frac{\pi}{2}\right)N  \right).
\end{align*}
Since the assumptions $\theta_N - \theta_0 < 0$ and $\kappa > 0$ give $m - n < 0$, we have
\begin{displaymath}
-(\theta_0 + \theta_N + \pi) n 
+ (\theta_N-\theta_0)l  + \left(\theta_0 + \frac{\pi}{2}\right)N \leq 0.
\end{displaymath}
Therefore we conclude
\begin{displaymath}
n \geq \max_{l=0,1,\ldots,n}
\frac{ \left(\theta_0 + \frac{\pi}{2}\right)N + (\theta_N-\theta_0)l } {\theta_N + \theta_0 + \pi}
=\frac{ \theta_0 + \frac{\pi}{2} }{\theta_N + \theta_0 + \pi}N,
\end{displaymath}
where we used the condition $l \geq 0$.  The remaining part of the estimate can be shown in a
similar manner. By the assumption \eqref{eqn:bdry_angle_assumption2}, we have
\begin{align*}
0 &\geq
\theta_0 - n  \kappa 
= \theta_0 - n \frac{\theta_N - \theta_0}{m - n}
= \frac{m \theta_0 - n \theta_N}{m - n}
= \frac{-n(\theta_0+\theta_N) + N\theta_0 }{m - n}.
\end{align*}
The condition $m - n < 0$ implies that the numerator of the last expression is non-negative, and
therefore we have
\begin{displaymath}
n \leq \frac{\theta_0}{\theta_0 + \theta_N}N,
\end{displaymath}
which proves the first case. The second case is proved in a similar manner noting that $m-n>0$. 
\end{proof}
Note that it is straightfoward to deduce that the indices $n$ of inflection satisfying
\eqref{eqn:angle_condition} exist if
\begin{equation}
 N\geq \frac{\pi}{2}\left|\frac{(\theta_0+\theta_N)(\theta_0+\theta_N+\pi)}{\theta_0-\theta_N}\right|.
\end{equation}

In summary, one can compute $(\delta,z)$ to generate dLAC with an inflection from the given data
$(N,a,\gamma_0,\gamma_{N+1},\theta_0,\theta_{N})$ for each $n$ in the relevant range
\eqref{eqn:position_estimate} as follows:
\begin{enumerate}
 \item Solve \eqref{eq:inflection_vertical} to obtain $\delta$. 
 \item Compute $z$ using \eqref{eq:inflection_horizontal} for given $\ell$.
\end{enumerate}
This computation generates several dLACs and the choice may be left to the user, but a criteria may
be given as follows. Consider the discrete fairing energy
\begin{equation}\label{eqn:discrete_fairing_functional2}
 \varPhi^{\lambda,a}(\gamma) = \sum_{k=0}^{N-1}
\left\{(u_k)^a + \frac{1}{(u_k)^a} + \lambda\left(\frac{q_{0}q_{N}}{q_kq_{k+1}}\right)^a\right\},
\end{equation}
whose Euler-Lagrange equation is given by
\begin{equation}\label{eqn:dE-L2}
 (u_{k})^a -   (u_{k-1})^a + \lambda \left(\frac{q_0q_{N}}{q_{k-1}q_{k}}\right)^a=0.
\end{equation}
%
\begin{prop}
The discrete fairing energy for dLAC with an inflection \eqref{eqn:dLAC_q} is given by
\begin{align}\label{eqn:discreteFairingEnergy_dLAC}
 \varPhi^{\lambda,a}(\gamma) &= \sum_{k=0}^{N-1}
= 2N + 2\left(\frac{2}{\delta} - \frac{1}{n+\delta}-\frac{1}{-n+N+\delta} \right).
\end{align}
\end{prop}
%
\begin{proof}
We first note that we have from \eqref{eqn:dLAC_q}
\begin{equation}\label{eqn:q_rel_dLAC}
 (q_{k+1})^a - (q_k)^a = \left\{
\begin{array}{cl}
-z^a& (k=0,1,\ldots,n-1),\\
z^a & (k=n,n+1,\ldots,N-1),
\end{array}\right.
\end{equation}
and that $(u_k)^a=(q_{k+1})^a/(q_k)^a$ $(k=0,1,\ldots N-1)$ satisfies
\begin{equation}\label{eqn:u_rel_dLAC}
(u_k)^a + \frac{1}{(u_{k-1})^a}=\left\{
\begin{array}{cl}\medskip
 {\displaystyle 2 }& (k\neq n), \\
 {\displaystyle \frac{2(1+\delta)}{\delta} }& (k=n).
\end{array}
 \right.
\end{equation}
The constant $\lambda$ in \eqref{eqn:dE-L2} for dLAC can be computed by using \eqref{eqn:q_rel_dLAC} as
\begin{align}
 \lambda & = - \Big((u_{k})^a -   (u_{k-1})^a\Big) \left(\frac{q_{k-1}q_{k}}{q_0q_{N}}\right)^a
=-\frac{(q_{k+1})^a(q_{k-1})^a-(q_k)^{2a}}{(q_{k-1})^a(q_k)^a}\ 
\left(\frac{q_{k-1}q_{k}}{q_0q_{N}}\right)^a\nonumber\\
& =-\frac{\big((q_{k})^a+z^a\big)\big((q_{k})^a-z^a\big)-(q_k)^{2a}}{(q_{k-1})^a(q_k)^a}\ 
\left(\frac{q_{k-1}q_{k}}{q_0q_{N}}\right)^a = \left(\frac{z^2}{q_0q_{N}}\right)^a. \label{eqn:dLAC_lambda}
\end{align}
Then we see by using \eqref{eqn:u_rel_dLAC} that
\begin{equation}\label{eqn:FE_1st}
 \sum_{k=0}^{N-1}\left((u_k)^a + \frac{1}{(u_k)^a}\right) = 2(N-2)+ (u_0)^a +2(u_n)^a + \frac{1}{(u_{N-1})^a},
\end{equation}
where we used $1/(u_{n-1})^a = (u_{n})^a$. We also have by using \eqref{eqn:q_rel_dLAC} and
\eqref{eqn:dLAC_lambda}
\begin{align}\label{eqn:FE_2nd}
&  \sum_{k=0}^{N-1}\lambda\left(\frac{q_{0}q_{N}}{q_kq_{k+1}}\right)^a
=  \sum_{k=0}^{N-1}\left(\frac{z^2}{q_kq_{k+1}}\right)^a
=  z^a\left(-\sum_{k=0}^{n-1}\frac{q_{k+1}^a - q_k^a}{(q_kq_{k+1})^a} 
   +\sum_{k=n}^{N-1}\frac{q_{k+1}^a - q_k^a}{(q_kq_{k+1})^a}\right)\nonumber\\
=& z^a\left(\frac{2}{(q_n)^a} - \frac{1}{(q_0)^a} - \frac{1}{(q_N)^a}\right).
\end{align}
Then, noticing by \eqref{eqn:q_rel_dLAC} that
\begin{equation}\label{eqn:FE_3rd}
\begin{split}
&(u_0)^a = \frac{(q_1)^a}{(q_0)^a} = \frac{(q_0)^a-z^a}{(q_0)^a}=1-\frac{z^a}{(q_0)^a},\\
&(u_n)^a = \frac{(q_{n+1})^a}{(q_n)^a} = 1+\frac{z^a}{(q_n)^a},\quad
\frac{1}{(u_{N-1})^a} = \frac{(q_{N-1})^a}{(q_{N})^a} = 1-\frac{z^a}{(q_{N})^a}, 
\end{split}
\end{equation}
we obtain from \eqref{eqn:FE_1st}, \eqref{eqn:FE_2nd} and \eqref{eqn:FE_3rd}
\begin{align}
 \varPhi^{\lambda,a}(\gamma) &= 
2(N-2)+ 1-\frac{z^a}{(q_0)^a} +2\left( 1+\frac{z^a}{(q_n)^a}\right) + 1-\frac{z^a}{(q_{N})^a}
+ z^a\left(\frac{2}{(q_n)^a} - \frac{1}{(q_0)^a} - \frac{1}{(q_N)^a}\right)\nonumber\\
& = 2N + 2z^a\left(\frac{2}{(q_n)^a} - \frac{1}{(q_0)^a} - \frac{1}{(q_N)^a}\right)
= 2N + 2\left(\frac{2}{\delta} - \frac{1}{n+\delta}-\frac{1}{-n+N+\delta} \right).
\end{align}
which proves the statement.
\end{proof}
%
We may choose the dLAC that attains the minimum of the discrete fairing energy
\eqref{eqn:discreteFairingEnergy_dLAC}.  Practically, we may choose the dLAC corresponding to the
maximum value of $\delta$ among those generated, since \eqref{eqn:discreteFairingEnergy_dLAC} is
monotonically decreasing with respect to $\delta>0$ for each $n$, and the change of the
energy with respect to $n$ is much smaller than that with respect to $\delta$.

\begin{center}
\begin{figure}[h]
\begin{center}
\includegraphics[width=1.0\linewidth]{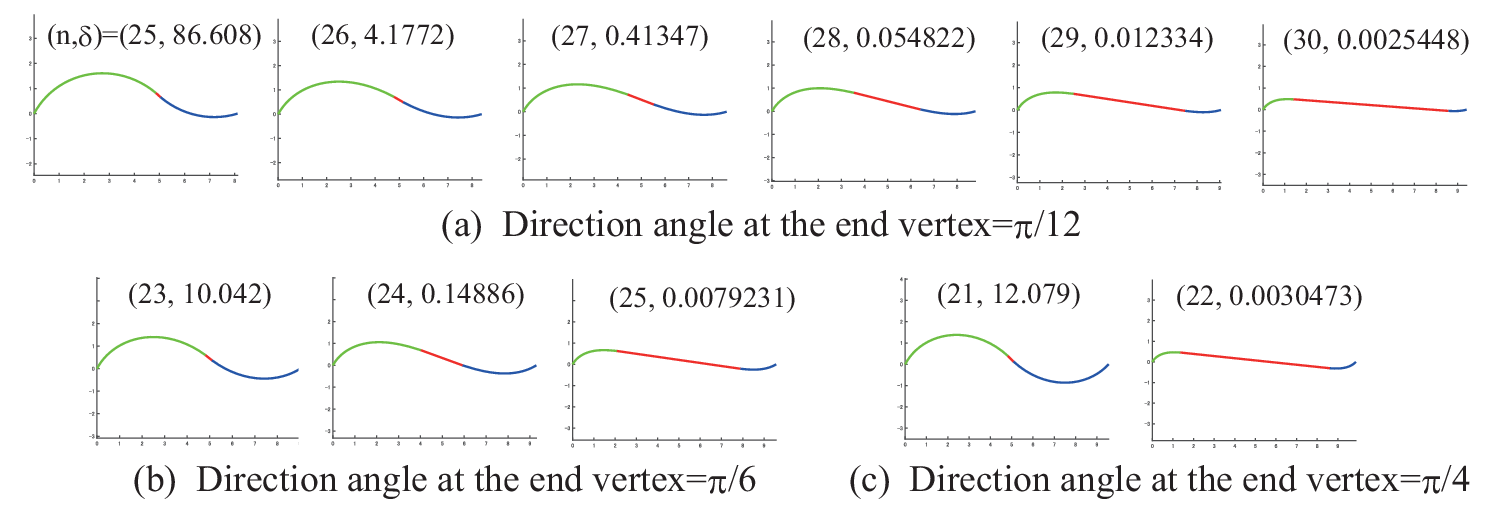} 
\end{center}
\caption{dLAC examples with $\alpha=-2/3$, $N=39$ and $\theta_{0}=\pi/3$}. \label{fig:dLACInflection}
\end{figure}
\end{center}

Figure \ref{fig:dLACInflection} shows various dLAC examples with an inflection edge. We specified
$N=39$, so the total number of the vertices is $41$. $\alpha=-2/3$ and the direction angle
$\theta_0$ at the start (left) vertex is equal to $\pi/3$. We changed the direction angle
$\theta_{39}$ at the end (right) vertex to be $\pi/12$, $\pi/6$ and $\pi/4$.  The edge in red is an
inflection edge in each curve. The sign of the discrete turning angle $\kappa$ of the curve segment
in green is negative and that in blue is positive.  For $\theta_{39}=\pi/12$, there are 6 solutions
for $(\delta,z)$, while for $\theta_{39} = \pi/6$ and $\theta_{39}=\pi/4$, $3$ and $2$ solutions
exist, respectively.  For each curve, we described its corresponding $(n,\delta)$ values. As $n$
increases, $\delta$ decreases.  The discrete fairing energy is the lowest for the left curve in each
group, in which its inflection edge is the shortest.
%
\begin{rem}
More rigourously, the summation of the fairing energy \eqref{eqn:discrete_fairing_functional2} for
the dLAC with an inflection should be taken separately for the left and right sides of the
inflection, respectively, and so the variation with the boundary conditions in Theorem
\ref{mainresult_discrete} imposed on both cases. The Euler-Lagrange equation \eqref{eqn:dE-L2}
covers the case of $k\neq n$ in \eqref{eqn:u_rel_dLAC}, but not the case of $k=n$. Nervertheless,
the computations in the above proof is valid, and the criterion of choosing the largest $\delta$
implies the smallest $q_n$, which seems natural.  This is an intricate problem and more analysis
needs to be done to make a rigorous statement. This problem will be dealt with in a future
publication.
\end{rem}
%

\section*{Acknowledgment}
This work was supported by JST CREST Grant Number JPMJCR1911. It was also supported by JSPS KAKENHI
Grant Numbers JP19K03461, JP19H02048, JP25289021, JP16H03941, \hfill\break JP16K13763, JP15K04834,
JP26630038, JST RISTEX Service Science, Solutions and Foundation Integrated Research Program, and
ImPACT Program of the Council for Science, Technology and Innovation. The authors acknowledge the
support by IMI Joint Use Program Short-Term Joint Research ``Differential Geometry and Discrete
Differential Geometry for Industrial Design'' (September 2016) and ``New developments of Discrete
Differential Geometry: from industrial design to architecture'' (September 2018). The authors would
like to express their sincere gratitude to Prof. Miyuki Koiso, Prof. Hiroyuki Ochiai, Prof. Nozomu
Matsuura and Prof. Sampei Hirose for invaluable comments and fruitful discussions.

\end{document}